\documentclass[11pt,twoside,a4paper,cmspaper,final,collab]{cms-tdr}
\def\svnVersion{64df8ca}\def\svnDate{2026/10/29}\def\cmsCernNoTag{CERN-EP-2026-126}\def\cmsCernDate{\today}\def\cmsMessage{Published in the Journal of High Energy Physics as \href{http://dx.doi.org/10.1007/JHEP10(2026)194}{\doi{10.1007/JHEP10(2026)194}.}}
\begin{document}\cmsNoteHeader{SMP-25-013}

\providecommand{\Pellpr}{{\HepParticle{\Pell}{}{\prime}}\xspace} 
\providecommand{\Pellmpr}{{\HepParticle{\Pell}{}{\prime-}}\xspace} 
\providecommand{\Pellppr}{{\HepParticle{\Pell}{}{\prime+}}\xspace} 
\providecommand{\Pellpmpr}{{\HepParticle{\Pell}{}{\prime\pm}}\xspace} 
\providecommand{\Pellmppr}{{\HepParticle{\Pell}{}{\prime\mp}}\xspace} 
\newcommand{\mll}{\ensuremath{m_{\Pell\Pell}}\xspace}
\newcommand{\mlll}{\ensuremath{m_{\Pell\Pell\Pell}}\xspace}
\newcommand{\jet}{\ensuremath{\mathrm{j}}}
\newcommand{\ptj}{\pt^\jet\xspace}
\newcommand{\mjj}{\ensuremath{m_{\jet\jet}}\xspace}
\newcommand{\mZ}{\ensuremath{m_{\PZ}}\xspace}
\newcommand{\detajj}{\ensuremath{\abs{\Delta\eta_{\jet\jet}}}\xspace}
\newcommand{\dphijj}{\ensuremath{\Delta\phi_{\jet\jet}}\xspace}
\newcommand{\WW}{\ensuremath{\PWpm\PWpm}\xspace}
\newcommand{\WZ}{\ensuremath{\PW\PZ}\xspace}
\newcommand{\tZq}{\ensuremath{\PQt\PZ\Pq}\xspace}
\newcommand{\tVx}{\ensuremath{\PQt\PV\mathrm{x}}\xspace}
\newcommand{\ptmax}{\ensuremath{\pt^{\text{max}}}\xspace}
\newcommand{\zepmax}{\ensuremath{\max(z_{\Pell}^{*})}\xspace}
\providecommand{\bigabs}[1]{\ensuremath{\Bigl|#1\Bigr|}\xspace}
\newcommand{\MocaPlus}{\textsc{MoCaNLO}+\textsc{Recola}\xspace}
\newcommand{\lumiTotal}{\ensuremath{171}\xspace}
\newcommand{\UParT}{{\textsc{UParT}}\xspace}
\newcommand{\NJ}{\ensuremath{n_{\mathrm{j}}}\xspace}

\newlength\cmsTabSkip\setlength{\cmsTabSkip}{1ex}
\providecommand{\cmsTable}[1]{\resizebox{\textwidth}{!}{#1}}
\providecommand{\cmsLeft}{left\xspace}
\providecommand{\cmsRight}{right\xspace}

\cmsNoteHeader{SMP-25-013}

\title{First measurements of vector boson scattering in \texorpdfstring{$\PWpm\PWpm$}{W+/- W+/-} and \texorpdfstring{$\PW\PZ$}{WZ} production in all-leptonic final states at \texorpdfstring{$\sqrt{s} = 13.6\TeV$}{sqrt(s) = 13.6 TeV}}

\date{\today}

\abstract{
The production cross sections of $\PWpm\PWpm$ and $\PW\PZ$ boson pairs in association with two jets in proton-proton collisions are measured at $\sqrt{s}=13.6\TeV$. The data sample corresponds to an integrated luminosity of 171\fbinv, collected with the CMS detector during 2022--2024. The measurements are performed in the leptonic decay modes: $\PWpm\PZ \to \Pellpm\PGn\Pellpmpr\Pellmppr$ and $\PWpm\PWpm \to \Pellpm\PGn\Pellpmpr\PGn$, where $\Pell, \Pellpr = \Pe$ or \PGm. The electroweak productions of $\PWpm\PWpm$ and $\PW\PZ$  bosons are each observed with a significance greater than five standard deviations from the background-only hypothesis. Differential fiducial cross sections as functions of several observables are also measured.
}

\hypersetup{%
pdfauthor={CMS Collaboration},%
pdftitle={First measurements of vector boson scattering in WW and WZ production in all-leptonic final states at sqrts = 13.6 TeV},%
pdfsubject={CMS},%
pdfkeywords={CMS, diboson, electroweak}
}

\maketitle

\section{Introduction}
\label{sec:Introduction}

\begin{figure*}[htb]
\centering
\includegraphics[width=0.49\textwidth]{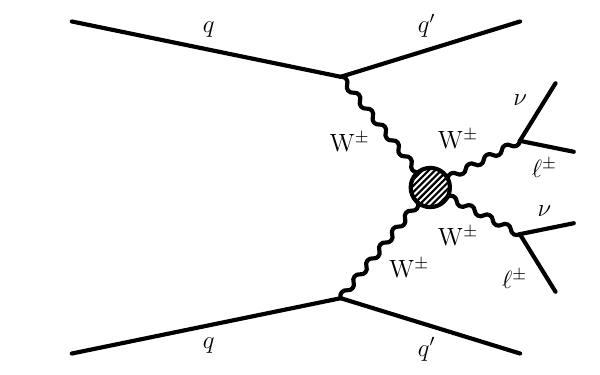}
\includegraphics[width=0.49\textwidth]{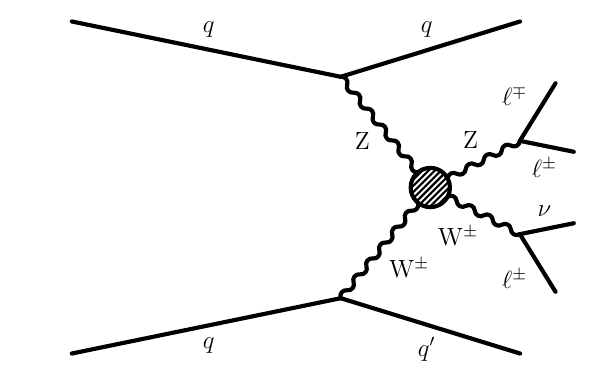}
\caption{Representative Feynman diagrams of a VBS process contributing to the EW-induced production of events containing $\WW$ (left) and $\WZ$ (right) boson pairs decaying to leptons, and two forward jets. New physics in the EW sector can modify the quartic gauge couplings, represented by a hatched circle.\label{fig:feynman}}
\end{figure*}

\begin{figure*}[htb]
\centering
\includegraphics[width=0.49\textwidth]{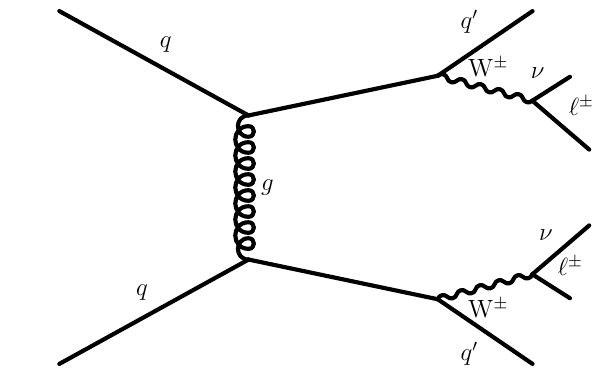}
\includegraphics[width=0.49\textwidth]{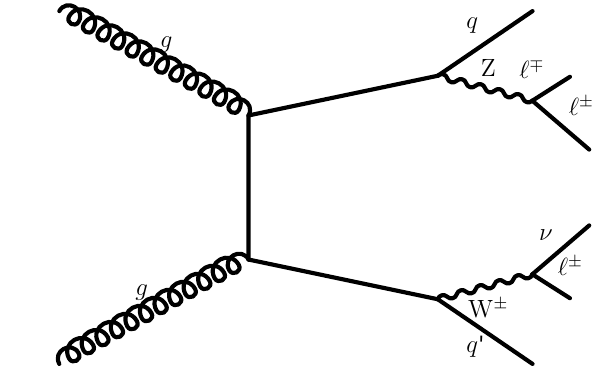}
\caption{Representative Feynman diagrams of the QCD-induced production of $\WW$ (left) and $\WZ$ (right) boson pairs decaying 
to leptons, and two jets.\label{fig:feynman_qcd}}
\end{figure*}

The observation of a Higgs boson with a mass of 125\GeV~\cite{AtlasPaperCombination,CMSPaperCombination,CMSPaperCombination2} has established the existence of the scalar field responsible for the electroweak (EW) symmetry-breaking mechanism, and measurements of its properties so far are consistent with the predictions of the standard model (SM) scalar sector. This includes the generation of masses for the \PW and \PZ gauge bosons via the Brout--Englert--Higgs mechanism~\cite{PhysRevLett.13.321,Higgs:1964ia,PhysRevLett.13.508,PhysRevLett.13.585,PhysRev.145.1156,PhysRev.155.1554}. Further insight into the electroweak symmetry breaking mechanism can be achieved through measurements of vector boson scattering (VBS) processes~\cite{Espriu:2012ih,Chang:2013aya}. At the CERN LHC, VBS interactions are characterized by the presence of two gauge bosons produced in association with two forward jets with large dijet invariant mass and rapidity separation, as shown in Fig.~\ref{fig:feynman}. They are part of a class of processes contributing to the diboson plus two jets production that proceeds via the EW interaction, referred to as EW-induced diboson production, which is $\mathcal{O}(\alpha^4)$ at tree level, where $\alpha$ is the charged EW coupling. An additional contribution to the diboson states arises via quantum chromodynamics (QCD) radiation of partons from incoming quarks or gluons, leading to tree-level contributions at $\mathcal{O}(\alpha^2\alpS^2)$, where $\alpS$ is the strong coupling. Figure~\ref{fig:feynman_qcd} shows representative Feynman diagrams for this class of processes, referred to as QCD-induced diboson production.

Modifications of the VBS production cross sections are predicted in models of physics beyond the SM, for example through changes to the Higgs boson couplings to gauge bosons~\cite{Espriu:2012ih,Chang:2013aya}. In addition, the non-Abelian gauge structure of the EW sector of the SM predicts self-interactions between gauge bosons through triple and quartic gauge couplings, which can be probed via measurements of VBS processes~\cite{Lee:1977yc,Lee:1977eg}. The possible presence of anomalous quartic gauge couplings would generally result in an excess of VBS events with respect to the SM predictions~\cite{Eboli:2006wa,kozow2019same,Kalinowski:2018oxd,Bellan:2021dcy,Ethier:2021ydt,Dedes:2020xmo,Fuks:2020att}.

The EW production of $\WW$ and $\WZ$ at the LHC in the leptonic decay modes has been previously studied at $\sqrt{s}=8$ and 13\TeV~\cite{Khachatryan:2014sta,Aad:2014zda,Sirunyan:2017ret,Aaboud:2019nmv,ATLAS:2023sua,CMS:2020gfh,Aad:2016ett,Sirunyan:2019ksz,Aaboud:2018ddq,ATLAS:2024ini,CMS:2021wlt}. The ATLAS and CMS Collaborations reported observations of the EW $\WW$ and $\WZ$ productions at 13\TeV with significances greater than five standard deviations from the background-only hypothesis. This paper presents a measurement of VBS in the $\WW$ and $\WZ$ channels using proton-proton ($\Pp\Pp$) collisions at $\sqrt{s}=13.6\TeV$. For the $\PW\PW$ measurement, the same-sign $\WW$ channel is studied because of its smaller background contributions compared to the $\PWpm\PWmp$ channel. The data sample corresponds to an integrated luminosity of $\lumiTotal\fbinv$ collected with the CMS detector in 2022, 2023, and 2024. The three data sets, corresponding to different LHC operating periods, are analyzed independently, with appropriate calibrations and corrections, to account for the various running conditions and detector performance.

The measurements are performed in the leptonic decay modes $\WW \to \Pellpm\PGn\Pellpmpr\PGn$ and $\PWpm\PZ\to\Pellpm\PGn\Pellpmpr\Pellmppr$, where $\Pell, \Pellpr = \Pe $ or $ \PGm$. Candidate events contain either two identified leptons of the same charge or three identified charged leptons with a total charge of $\pm1$, moderate missing transverse momentum ($\ptmiss$), and two jets with a large pseudorapidity separation ($\detajj$) and a large dijet mass ($\mjj$). The requirements on $\detajj$ and $\mjj$ reduce the contribution from the QCD-induced production of boson pairs in association with two jets. This makes the targeted experimental signature an ideal topology for studying VBS. The EW $\WW$ and $\WZ$ production cross sections are measured simultaneously by performing a binned maximum-likelihood fit of several distributions sensitive to these processes. Tabulated results are provided in the HEPData record for this work~\cite{HEPData}.

\section{The CMS detector}
\label{sec:cms}

The CMS apparatus~\cite{CMS:2008xjf,CMS:2023gfb} is a multipurpose, nearly hermetic detector, designed to trigger on~\cite{CMS:2020cmk,CMS:2016ngn,CMS:2024aqx} and identify electrons, muons, photons, and (charged and neutral) hadrons~\cite{CMS:2020uim,CMS:2018rym,CMS:2014pgm}. Its central feature is a superconducting solenoid of 6\unit{m} internal diameter, providing a magnetic field of 3.8\unit{T}. Within the solenoid volume are a silicon pixel and strip tracker, a lead tungstate crystal electromagnetic calorimeter (ECAL), and a brass and scintillator hadron calorimeter (HCAL), each composed of a barrel and two endcap sections. Forward calorimeters extend the $\eta$ coverage provided by the barrel and endcap detectors. Muons are reconstructed using gas-ionization detectors embedded in the steel flux-return yoke outside the solenoid. More detailed descriptions of the CMS detector, together with a definition of the coordinate system used and the relevant kinematic variables, can be found in Refs.~\cite{CMS:2008xjf,CMS:2023gfb}. 

Events of interest are selected using a two-tiered trigger system. The first level, composed of custom hardware processors, uses information from the calorimeters and muon detectors to select events at a rate of around 100\unit{kHz} within a fixed latency of 4\mus~\cite{CMS:2020cmk}. The second level, known as the high-level trigger, consists of a farm of processors running a version of the full event reconstruction software optimized for fast processing, and reduces the event rate to a few kHz before data storage~\cite{CMS:2016ngn,CMS:2024aqx}.

\section{Signal and background simulation}
\label{sec:samples}

Several Monte Carlo (MC) event generators are used to simulate the signal and background processes. The simulated samples are used to optimize the event selection, evaluate selection efficiencies and systematic uncertainties, and compute expected yields. The SM EW $\WW$ and $\WZ$ processes, where both bosons decay leptonically, are simulated using the event generator \MGvATNLO 2.9.9~\cite{Alwall:2014hca,MadGraph,Frederix2012} at leading order (LO) accuracy with six EW ($\mathcal{O}(\alpha^6)$) and zero QCD vertices. The \MGvATNLO generator is also used to simulate at LO the QCD-induced $\WW$ process. Contributions with an initial-state {\cPqb} quark are excluded from the EW $\WZ$ simulation because they are considered part of the $\tZq$ background process. Triboson processes, where a $\WZ$ boson pair is accompanied by a third vector boson that decays into jets, are included using \MGvATNLO. The QCD-induced $\WZ$ process is simulated at LO with up to three additional partons in the matrix element calculations using the \MGvATNLO generator with at least one QCD vertex at tree level. The different jet multiplicities are merged using the MLM scheme~\cite{MLMmerging} to match matrix element and parton shower jets, and the inclusive contribution is normalized to next-to-next-to-LO (NNLO) predictions~\cite{Grazzini:2016swo}.

A complete set of next-to-LO (NLO) QCD and EW corrections for the leptonic $\WW$ scattering process has been computed~\cite{Biedermann:2016yds,Biedermann:2017bss}, and they are used in this study. They reduce the LO cross section of the EW $\WW$ process by 10--15\%, with the correction increasing in magnitude with the dilepton and dijet invariant masses. Similarly, for the leptonic $\WZ$ scattering process, the NLO QCD and EW corrections, computed at orders $\mathcal{O}(\alpS\alpha^6)$ and $\mathcal{O}(\alpha^7)$~\cite{Denner:2019tmn,ParticleDataGroup:2024cfk}, reduce the cross sections for the EW $\WZ$ process by about 10\%. The interference between the EW and QCD diagrams is also produced with \MGvATNLO. The contribution of the interference is considered to be part of the QCD production, leading to an increase of about 5\% of the predicted QCD processes in the fiducial region.

The $\POWHEG$ v2~\cite{Frixione:2002ik,Nason:2004rx,Frixione:2007vw,Alioli:2008gx,Alioli:2010xd} generator is used to simulate the $\ttbar$ and $\PQt\PW$ processes at NLO accuracy for the inclusive production. The production of $\ttbar\PW$, $\ttbar\PZ$, $\ttbar\PGg$, and triple vector boson ($\PV\PV\PV$) background events is simulated at NLO accuracy in QCD using the \MGvATNLO generator~\cite{Alwall:2014hca,MadGraph,Frederix2012}. The $\tZq$ process is simulated at NLO in the four-flavor scheme using \MGvATNLO. This MC simulation is normalized using a cross section computed at NLO with \MGvATNLO in the five-flavor scheme, following the procedure of Ref.~\cite{Sirunyan:2017nbr}. The double parton scattering (DPS) $\WW$ production is generated at LO using $\PYTHIA$ 8.306~\cite{Bierlich:2022pfr}. A summary of the main simulated samples used in the analysis is shown in Table~\ref{tab:MCsamples}.

The NNPDF 3.1 NNLO~\cite{Ball:2017nwa} parton distribution functions (PDFs) are used for simulating all samples. The underlying event (UE) parton showering and hadronization are simulated using $\PYTHIA$. The UE modeling is generated using the CP5~\cite{Sirunyan:2019dfx} tune. All MC events are processed through a simulation of the CMS detector based on \GEANTfour~\cite{Agostinelli:2002hh} and are reconstructed with the same algorithms used for data. Additional $\Pp\Pp$ interactions in the same and nearby bunch crossings, referred to as pileup, are also simulated. The distribution of the number of pileup interactions in the simulation is adjusted to match the one observed in the data, with an average pileup of about 50 interactions per bunch crossing.

\begin{table*}[htbp]
\centering
  \topcaption{Summary of the main simulated samples used in the analysis.\label{tab:MCsamples}}
\begin{tabular}{lcc}
\hline
Process & Matrix-element generator & Accuracy \\
\hline
EW $\WW$                               & \MGvATNLO 2.9.9 & LO  \\
EW $\WZ$                               & \MGvATNLO 2.9.9 & LO  \\
QCD $\WW$                              & \MGvATNLO 2.9.9 & LO  \\
QCD $\WZ$                              & \MGvATNLO 2.9.9 & LO  \\
EW--QCD interference                   & \MGvATNLO 2.9.9 & LO  \\
$\ttbar$, $\PQt\PW$                    & \POWHEG\ v2     & NLO \\
$\ttbar\PW$, $\ttbar\PZ$, $\ttbar\PGg$ & \MGvATNLO 2.9.9 & NLO \\
$\PV\PV\PV$                            & \MGvATNLO 2.9.9 & NLO \\
$\PQt\PZ\PQq$                          & \MGvATNLO 2.9.9 & NLO \\
DPS $\WW$                              & \PYTHIA\ 8.306  & LO  \\
\hline
\end{tabular}
\end{table*}

\section{Event reconstruction}
\label{sec:objects}

A particle-flow (PF) algorithm~\cite{CMS:2017yfk} aims to reconstruct and identify each individual particle in an event, with an optimized combination of information from the various elements of the CMS detector. The energy of photons is obtained from the ECAL measurement. The energy of an electron is determined from a combination of the electron momentum at the primary interaction vertex as determined by the tracker, the energy of the corresponding ECAL cluster, and the energy sum of all bremsstrahlung photons spatially compatible with originating from the electron track. The energy of a muon is obtained from the curvature of the corresponding track. The energy of a charged hadron is determined from a combination of the momentum measured in the tracker and the matching ECAL and HCAL energy deposits, corrected for the response function of the calorimeters to hadronic showers. Finally, the energy of neutral hadrons is obtained from the corresponding corrected ECAL and HCAL energies. The primary vertex (PV) is taken to be the vertex corresponding to the hardest scattering in the event, evaluated using tracking information alone, as described in Section 9.4.1 of Ref.~\cite{CMS-TDR-15-02}.

For each event, hadronic jets are clustered from these reconstructed particles using the anti-\kt algorithm~\cite{Cacciari:2008gp,Cacciari:2011ma} with a distance parameter of 0.4. The jet momentum is determined as the vectorial sum of all particle momenta in the jet, and is found from simulation to differ less than 5--10\%, on average, from the true momentum over the entire transverse momentum (\pt) spectrum and detector acceptance. Jet energy corrections are derived from simulation to bring, on average, the measured response of jets close to that of generator-level jets. In situ measurements of the momentum balance in dijet, $\PGg${+}jet, \PZ{+}jet, and multijet events are used to account for any residual differences in the jet energy scale (JES) between data and simulation~\cite{CMS:2016lmd}. The jet energy resolution (JER) amounts typically to 15--20\% at 30\GeV, 10\% at 100\GeV, and 5\% at 1\TeV~\cite{CMS:2016lmd}. Selection criteria are applied to remove jets potentially dominated by anomalous contributions from various subdetector components or reconstruction failures~\cite{Sirunyan:2019kia}. 

To mitigate the effect of pileup interactions at the reconstructed particle level, the pileup-per-particle identification algorithm (PUPPI)~\cite{CMS:2020ebo,Bertolini:2014bba} is used.  A local shape variable is defined to identify particles that originate from hard scatter.  Charged particles from pileup vertices are discarded. For each neutral particle, a local shape variable is calculated using surrounding charged particles compatible with the PV within the tracker acceptance ($\abs{\eta} < 2.5$), and both charged and neutral particles outside of the tracker coverage. The momenta of neutral particles are rescaled based on their probability of originating from the PV, eliminating the need for jet-based pileup corrections~\cite{CMS:2020ebo}.

For this analysis, VBS jets are selected by requiring $\ptj>50\GeV$ and $\abs{\eta} < 4.7$. The jet counting multiplicity $\NJ$ is determined from the number of reconstructed jets with $\pt>30\GeV$ and $\abs{\eta} < 4.7$. Jets falling within $\Delta R = \sqrt{\smash[b]{(\Delta\eta)^2+(\Delta\phi)^2}} < 0.4$ (where $\phi$ is the azimuthal angle in radians) of a "loose" charged lepton, selected as described below, are discarded.

Jets with $\pt>20\GeV$ and $\abs{\eta}<2.5$ originating from \PQb quarks are identified (``tagged") by a multivariate algorithm based on the transformer architecture called \UParT~\cite{Qu:2022mxj,CMS-DP-2025-081}. For the chosen working point, the efficiency to select \PQb\ jets is about 90\% and the rate for incorrectly tagging jets originating from the hadronization of gluons or $\cPqu$, $\cPqd$, $\cPqs$ quarks is about 10\%. The rate for mistagging jets originating from the hadronization of $\cPqc$ quarks is about 50\%.
 
The \ptvecmiss is computed as the negative vector $\pt$ sum of all the PF candidates in an event, and its magnitude is denoted as \ptmiss~\cite{Sirunyan:2019kia}. The \ptvecmiss is corrected to account for corrections to the energy scale of the reconstructed jets in the event. The PUPPI algorithm is applied to reduce the pileup dependence of the \ptvecmiss observable. The \ptvecmiss is computed from the PF candidates weighted by their probability to originate from the PV~\cite{Sirunyan:2019kia}. 

Electrons and muons are reconstructed by associating a track reconstructed in the tracking detectors with either a cluster of energy in the ECAL~\cite{Khachatryan:2015hwa} or a track in the muon system~\cite{CMS:2018rym}. Several levels of lepton selection are employed in this analysis, following standard practice in leptonic measurements, with the definitions provided in Refs.~\cite{Khachatryan:2015hwa,CMS:2018rym,Sirunyan:2018egh}. A very loose preselection, including basic kinematic requirements such as $\pt$ and $\eta$, is used to define the lepton multiplicity categories (two- and three-lepton events). Electrons (muons) must have $\pt>10\GeV$ and $\abs{\eta}<2.5$ (2.4). A tighter set of identification and isolation criteria is then applied to define the final signal regions (SRs). In addition, an intermediate (``loose") selection, less restrictive than the tight one but more selective than the preselection, is used for the estimation of nonprompt-lepton backgrounds. In all cases, the selections involve not only kinematic requirements but also identification, isolation, and quality criteria. For simplicity, these are referred to as loose and tight working points throughout the text. To reduce the misidentification rates of nonprompt leptons, a multivariate discriminant~\cite{CMS:2018fdh} is applied. For electrons, the background contribution coming from a mismeasurement of the track charge is not negligible. The charge sign is evaluated with three methods determining the electron curvature; requiring all three charge evaluations to agree reduces this background contribution by a factor of five, with an efficiency of about 97\%~\cite{Khachatryan:2015hwa}. For muons, the charge mismeasurement is negligible~\cite{Chatrchyan:2009ae,Sirunyan:2019yvv}. 

\section{Event selection}
\label{sec:selection}

Collision events are collected using single-electron and single-muon triggers that require the presence of an isolated lepton with $\pt$ larger than 27 and 24\GeV, respectively. In addition, a set of dilepton triggers with lower $\pt$ thresholds are used, ensuring a trigger efficiency above 99\% for events that satisfy the subsequent offline selection.

Several selection requirements are used to isolate the VBS topology by reducing the contributions from background processes. Background-enriched control regions (CRs) are built by inverting some of these selection requirements. In the offline analysis, events with two or three isolated charged leptons with $\pt>10\GeV$ and at least two jets with $\pt^\jet>50\GeV$ and $\abs{\eta}<4.7$ are accepted as candidate events. Candidate events with at least four (three) loose charged leptons are rejected from the $\WZ$ ($\WW$) channel. To reduce top quark background contributions, events with one or more \PQb-tagged jets are rejected.

In the $\WZ$ channel, the oppositely charged, same-flavor leptons forming the \PZ boson candidate are required to have $\pt>25$ (10)\GeV for the leading (subleading) lepton, with the invariant mass of the dilepton pair $\mll$ satisfying $\abs{\mll - \mZ}<15\GeV$. In candidate events with three same-flavor leptons, the oppositely charged lepton pair with the $\mll$ closest to $\mZ$~\cite{ParticleDataGroup:2024cfk} is selected as the \PZ boson candidate. The third lepton, selected with $\pt>20\GeV$, is associated with the \PW boson. In addition, the trilepton invariant mass, $\mlll$, is required to exceed 100\GeV.   

In the $\WW$ channel, the leading (subleading) lepton is required to have $\pt>25$ (20)\GeV, and $\mll$ must be greater than 20\GeV. Candidate events in the dielectron final state satisfying $\abs{\mll-\mZ}<15\GeV$ are rejected to reduce the number of \PZ boson background events where the charge of one of the electron candidates is misidentified. 

The VBS topology is targeted by requiring $\mjj>500\GeV$ and $\detajj>2.5$. The candidate $\WW$ ($\WZ$) events are also required to have $\zepmax<0.75$ (1.0), where 
\begin{equation}
z^{*}_{\Pell}=\bigabs{\eta^{\Pell} - \frac{\eta^{\jet_{1}} + \eta^{\jet_{2}}}{2}}/\detajj
\label{eq:zep}
\end{equation}
is the Zeppenfeld variable~\cite{Rainwater:1996ud}, $\eta^{\Pell}$ is the lepton pseudorapidity, and $\eta^{\jet_{1}}$ and $\eta^{\jet_{2}}$ are the pseudorapidities of the two candidate VBS jets. In the case of more than two jet candidates, the two jets with the largest \pt are selected. The discriminating power of the kinematic variables used in VBS event selection has been well established~\cite{Rainwater:1996ud}, and is illustrated in Ref.~\cite{CMS:2020gfh}. The Zeppenfeld variable measures the position of a particle (typically a boson or decay product) in rapidity relative to the midpoint of the two tagging jets. In VBS production, the bosons tend to be produced centrally between widely separated forward jets, whereas in QCD production they are more uniformly distributed, making this variable a powerful discriminator.

The $\ptmiss$, associated with the undetected neutrinos, is required to be greater than 30\GeV. The selection requirements defining the $\WW$ and $\WZ$ SRs are summarized in Table~\ref{tab:selectioncutsSR}. The $\WW$ SR is dominated by the EW signal process, whereas the $\WZ$ SR has large contributions from the QCD $\WZ$ process. The EW to QCD ratio is roughly 10 to 1 (1 to 3) for the $\WW$ ($\WZ$) production in the SR.

\begin{table*}[htbp]
\centering
  \topcaption{Selection criteria defining the $\WW$ and $\WZ$ SRs. The third lepton in the $\WZ$ selection refers to the single lepton from the decay of the \PW boson. The $\abs{\mll - \mZ}$ requirement is applied to the dielectron final state only in the $\WW$ SR.\label{tab:selectioncutsSR}}
\begin{tabular} {lcc}
\hline
  Variable & $\WW$ SR & $\WZ$ SR \\
\hline
Leptons                 & 2 leptons, $\pt>25/20\GeV$ & 3 leptons, $\pt>25/10/20\GeV$ \\
$\abs{\mll - \mZ}$      & $>$20\GeV ($\Pe\Pe$)       & $<$15\GeV		     \\
$\mll$                  & $>$20\GeV                  & \NA			     \\
$\mlll$                 & \NA                        & $>$100\GeV		     \\
$\ptmiss$               & \multicolumn{2}{c}{$>$30\GeV}                              \\
\PQb\ quark veto        & \multicolumn{2}{c}{Required}                               \\
$\pt^\jet$              & \multicolumn{2}{c}{$>$50\GeV}                              \\
$\zepmax$               & $<0.75$		     & $<$1.0  		             \\
$\mjj$                  & \multicolumn{2}{c}{$>$500\GeV}                             \\
$\detajj$               & \multicolumn{2}{c}{$>$2.5}                                 \\
\hline
\end{tabular}
\end{table*}

Two CRs, enriched in either nonprompt-lepton or $\tZq$ backgrounds, are built to estimate these processes from data. The nonprompt-lepton CR is defined by requiring the same selection as the $\WW$ SR, but with the \PQb\ jet veto requirement inverted. The selected events are enriched in nonprompt-lepton background events coming mostly from semileptonic $\ttbar$ events. Similarly, the $\tZq$ CR is defined by applying the same selection as the $\WZ$ SR, but with the \PQb\ jet veto requirement inverted. The selected events are dominated by the $\tZq$ background process.

The $\WZ$ SR is dominated by QCD $\WZ$ events after the kinematic selection described above. To enhance the relative contribution of EW $\WZ$ production, a multivariate analysis (MVA) discriminator is used. A Gradient Boosted Decision Tree (BDT) is trained using the \textsc{tmva} package~\cite{Hocker:2007ht}. The BDT is trained using EW $\WZ$ samples as signal and QCD $\WZ$ samples as background. An MVA discriminator has been developed that is largely independent of the specific diboson final state, as it utilizes a set of variables that can be constructed from any vector boson final state. Table~\ref{tab:BDT_variables} describes the list of input variables used in the BDT training.

\begin{table*}[htbp]
\centering
\topcaption{List and description of the input variables to the BDT used in the $\WZ$ SR. The Zeppenfeld variable is defined in Eq.~(\ref{eq:zep}), where $\eta$ refers either to a single lepton or to the trilepton system. The "all boson" object is built using the charged leptons and \ptvecmiss.\label{tab:BDT_variables}}

\begin{tabular}{ l c}
\hline
Variable & Definition \\
\hline
$\mjj$                            & Mass of the leading and trailing jets system \\
$\pt^{\jet\jet}$                  & \pt of dijet system \\
$\detajj$                         & Absolute difference in rapidity between the leading and trailing jets \\
$\dphijj$                         & Difference in $\phi$ between the leading and trailing jets \\
$\pt^{\jet1}$                     & \pt of the leading jet \\
$\pt^{\jet2}$                     & \pt of the trailing jet \\
$\eta^{\jet1}$                    & $\eta$ of the leading jet \\
$\eta^{\jet2}$                    & $\eta$ of the trailing jet \\
$\zepmax$                         & Maximum Zeppenfeld variable \\
$z_{3\Pell}^{*}$                   & Zeppenfeld variable of the trilepton system \\
$\pt^{\PV\PV}$                    & \pt of the vectorial sum of all boson objects \\
$\pt^{\text{tot}}$              & \pt of the vectorial sum of all boson objects and VBF jets \\
$\Delta \eta_{\jet1,\PV\PV}$      & $\Delta \eta$ between the leading jet and the trilepton system \\
$\Delta \eta_{\jet2,\PV\PV}$      & $\Delta \eta$ between the trailing jet and the trilepton system \\
\multirow{2}{*}{$\abs{\vec{\pt}^{\mathrm{tot}}}/\sum_i \pt^i$}    & Transverse component of the vector sum of the bosons \\
                                                                  &  and tagged jets momenta, normalized to their scalar \pt sum \\
\hline
\end{tabular}
\end{table*}

\section{Background estimation}
\label{sec:backgrounds}

A combination of methods based on CRs in data and simulation is used to estimate the background contributions. Uncertainties related to the theoretical and experimental predictions are estimated as described in Section~\ref{sec:systematics}. The normalization of the $\WZ$ process in the $\WW$ SR is constrained using data in the $\WZ$ SR, which is simultaneously included in the maximum likelihood fit used to extract the results, as described in Section~\ref{sec:results}. The background contribution from electron charge misidentification is estimated from simulation and data-to-simulation efficiency corrections are applied. The electron charge misidentification rate, estimated using Drell--Yan events, is about 0.01 (0.30)\% in the barrel (endcap) region~\cite{Khachatryan:2015hwa,CMS-DP-2018-017}.

The remaining contribution from the nonprompt-lepton background is estimated directly from data following the technique described in Ref.~\cite{Khachatryan:2014sta}. A ``pass--fail'' control sample is constructed by selecting events with two leptons: one that satisfies the tight lepton selection requirements, and another that does not meet the tight criteria but fulfills a looser set of requirements. The jet-to-lepton misidentification probability, denoted as $f$, corresponds to the fraction of objects passing the loose lepton selection that also satisfy the tight criteria. This probability is measured as a function of the lepton \pt and $\eta$ in a data sample enriched in QCD multijet events, which is dominated by nonprompt leptons. Event weights of $f/(1-f)$ are applied to the events in the pass--fail sample to estimate the nonprompt-lepton contribution in the SR~\cite{Sirunyan:2018egh}. The dominant systematic uncertainty associated with this background estimation method arises from the uncertainty in $f$. This uncertainty has two main components: the dependence of $f$ on the sample composition, and the methodological approach used to extract it. The first component is evaluated by varying the jet $\pt$ threshold in the QCD multijet sample and requiring at least one b-tagged jet, thereby altering the sample composition. The methodological uncertainty is assessed through a closure test where $f$ is extracted from and applied to simulated samples. Finally, the limited size of the control sample is also included as a source of the uncertainty.

The two CRs defined in Section~\ref{sec:selection} are used to estimate the normalization of the main background processes from data. The background estimation is further cross-checked in five validation regions (VRs). In four of these VRs, the requirements on $z^*_\Pell$, $\mjj$, and $\Delta\eta_{jj}$ are relaxed, the $\cPqb$-quark veto is either kept or inverted, and the remaining criteria are the same as in the $\WW$ and $\WZ$ SRs. Additionally, a $\PZ\PZ$ VR is defined identically to the $\WZ$ SR, with the additional requirement of four leptons forming two pairs with $\abs{\mll - \mZ} < 15\GeV$ each, and the relaxation of the $\ptmiss$ and $\cPqb$-quark veto requirements. 

\section{Systematic uncertainties}
\label{sec:systematics}

Multiple sources of systematic uncertainty are considered in the measurement of the inclusive and differential cross sections. The uncertainties are included through nuisance parameters in the profile likelihood fit~\cite{CMS:2024onh,Cowan:2010js}, described in Section~\ref{sec:results}. They can affect either the normalization or both the normalization and the shape of the distributions of the signal or background processes. For each source, the predicted event samples are used to construct histograms that describe the expected signal and background distributions for a specific nuisance parameter variation. In the fit of the templates to the data, the best fit values of the measured cross section and the nuisance parameters are determined simultaneously.

The uncertainty in the integrated luminosity measurement is 1.4, 1.3, and 2.0\%~\cite{CMS-PAS-LUM-22-001}, for 2022, 2023, and 2024, respectively. They are treated as partially correlated across the three data sets. The uncertainty in the simulation of pileup events is evaluated by varying the total inelastic $\Pp\Pp$ cross section of 69.2\unit{mb} by $\pm5\%$, which has an impact in the expected signal and background yields of about 1\%.

Discrepancies in the lepton reconstruction and identification efficiencies between data and simulation are corrected by applying scale factors to all simulation samples. These scale factors, which depend on the lepton \pt and $\eta$, are determined using $\PZ \to \Pell\Pell$ events recorded with independent triggers~\cite{CMS:2018rym,CMS:2020uim,Sirunyan:2019bzr}, with associated uncertainties of about 0.5\% for both electrons and muons. The uncertainty in the trigger efficiency determination leads to an uncertainty smaller than 1\% in the expected signal yield. The lepton momentum scale uncertainty is computed by varying the momenta of the leptons in simulation by their uncertainties, and repeating the analysis selection. The resulting uncertainties in the yields are about 0.1\% for both electrons and muons.

The uncertainty in the calibration of the JES and JER affects $\ptmiss$ and the jet multiplicity. These effects are estimated by changing the jet energy in the simulation up and down by one standard deviation. The uncertainty in the JES and JER is 2--5\%, depending on \pt and $\eta$~\cite{CMS:2016lmd}, and the impact on the expected signal and background yields is 3\%. There is a larger effect of the JES uncertainty in the EW $\WZ$ cross section measurement since the MVA discriminator is driven by jet-related variables, as seen in Table~\ref{tab:BDT_variables}.

The \cPqb\ tagging efficiency in the simulation is corrected using scale factors determined from data~\cite{Sirunyan:2017ezt}. These values are estimated separately for correctly and incorrectly identified jets. Each set of values results in uncertainties in the \cPqb\ tagging efficiency of about 1--4\%, and the impact on the expected signal and background yields is about 1\%. The uncertainties in the JES and \cPqb\ tagging are treated as partially correlated across the three data sets.

Because of the choice of the renormalization and factorization scales, the theoretical uncertainties are estimated by varying these scales independently up and down by a factor of two from their nominal values  (excluding the two extreme variations) and taking the largest cross section variations as the uncertainty~\cite{Ballestrero:2018anz}. The PDF uncertainties are evaluated according to the procedure described in Ref.~\cite{PDFLHC}. The statistical uncertainties that are associated with the limited number of simulated and data events used to estimate the nonprompt-lepton background are also considered as systematic uncertainties.

A summary of the relative systematic uncertainties in the EW $\WW$ and $\WZ$ cross sections is shown in Table~\ref{tab:ewSystematics}. The slightly larger theoretical and jet experimental uncertainties in the EW $\WZ$ cross section measurement arise from the difficulty in disentangling the EW and QCD components.

\begin{table*}[htbp]
\centering
\topcaption{
Uncertainties in the signal strength $\mu$, defined as the measured values of the inclusive production cross sections divided by the predicted cross sections. Lepton experimental uncertainties encompass the effects of the calibration of lepton momentum scale and resolution, as well as lepton trigger, reconstruction, identification, and isolation efficiencies. Jet experimental uncertainties encompass the effects of JES and JER. The limited sample size refers to the finite number of simulated and data events used to estimate the backgrounds.
\label{tab:ewSystematics}}
\begin{tabular}{lcc}
\hline
Source of uncertainty    & EW $\WW$ & EW $\WZ$ \\
\hline
Integrated luminosity    & 0.013 & 0.013 \\
Lepton experimental      & 0.016 & 0.030 \\
Jet experimental         & 0.023 & 0.061 \\
$\Pb$ tagging            & 0.016 & 0.004 \\
Nonprompt background     & 0.008 & 0.006 \\
Limited sample size      & 0.033 & 0.034 \\
Theory                   & 0.016 & 0.038 \\
Statistical              & 0.095 & 0.188 \\
\hline
Total                    & 0.107 & 0.204 \\
\hline
\end{tabular}
\end{table*}

\section{Results}
\label{sec:results}

To discriminate between the signals and the remaining backgrounds, a binned maximum-likelihood fit is performed using the $\WW$ and $\WZ$ SRs, and the nonprompt-lepton and $\tZq$ CRs, using the \texttt{combine} statistical analysis tool~\cite{CMS:2024onh}, based on the \textsc{RooFit}~\cite{Verkerke:2003ir} and \textsc{RooStats}~\cite{Moneta:2010pm} frameworks. The normalization factors for the nonprompt-lepton and $\tZq$ background processes are included as freely floating parameters in the maximum likelihood fit, together with the EW $\WW$, EW $\WZ$, and QCD $\WZ$ cross sections. For the EW $\WW$ cross section measurements, the QCD $\WW$ contribution is small and taken from the SM prediction. The systematic uncertainties are treated as nuisance parameters in the fit~\cite{Junk,Read}.  

The $\mjj$ variable is effective in discriminating between the signal and background processes because VBS topologies typically exhibit larger $\mjj$. The $\mll$ variable is also effective in discriminating between signal and background processes because the nonprompt-lepton processes tend to have smaller $\mll$. A two-dimensional distribution is used in the fit of the $\WW$ SR, with four bins in $\mjj$ ([500, 900, 1300, 2000, $\infty$]\GeV) and four bins in $\mll$  ([20, 80, 140, 220, $\infty$]\GeV). The BDT score distribution is used as the observable in the $\WZ$ SR, which is split into two categories with $z_{3\Pell}^{*}$ smaller or larger than 0.25. The $\mjj$ distribution is used as observable in the CRs, using four bins ([500, 900, 1300, 2000, $\infty$]\GeV).

\begin{figure*}[htbp]
\centering
\includegraphics[width=0.41\textwidth]{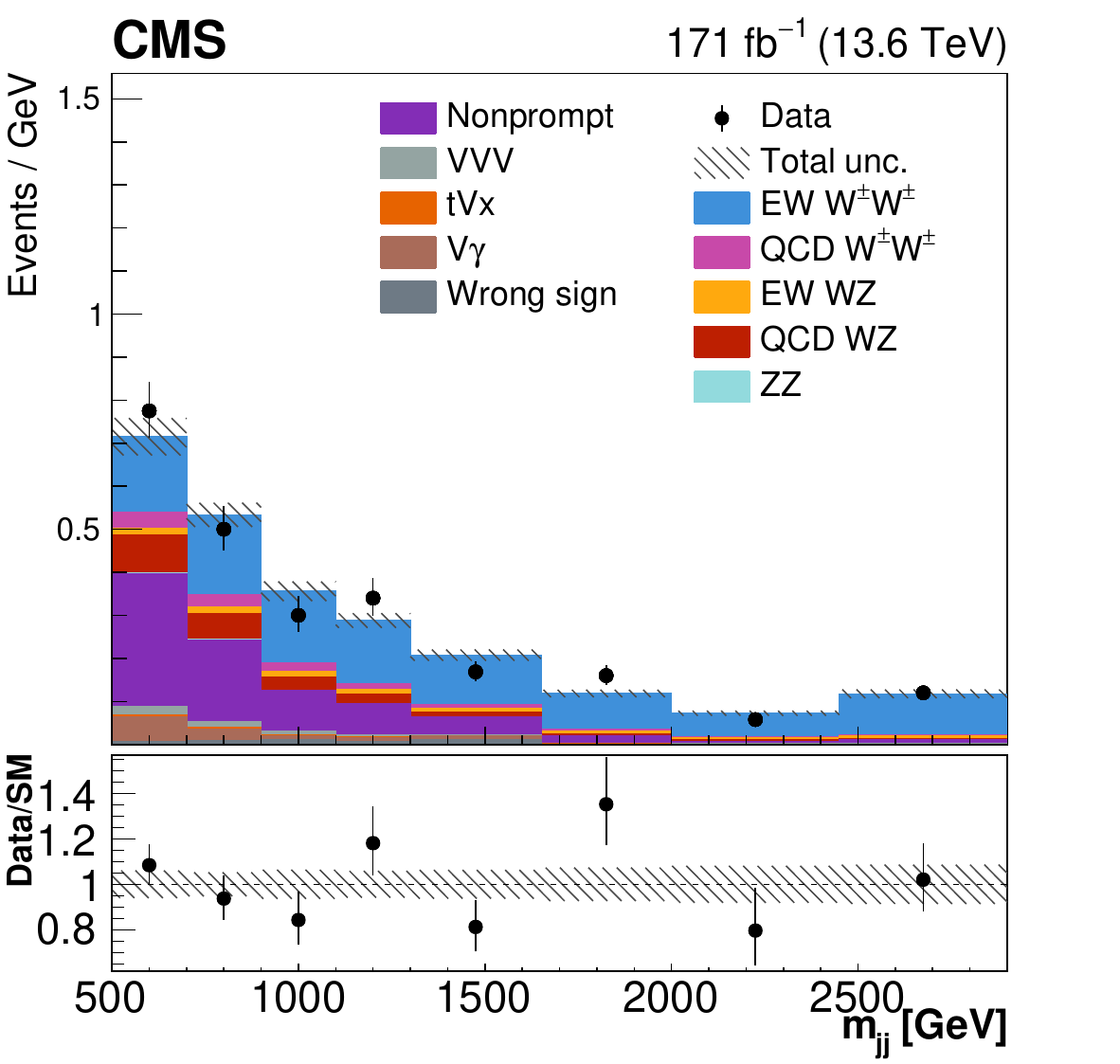}
\includegraphics[width=0.41\textwidth]{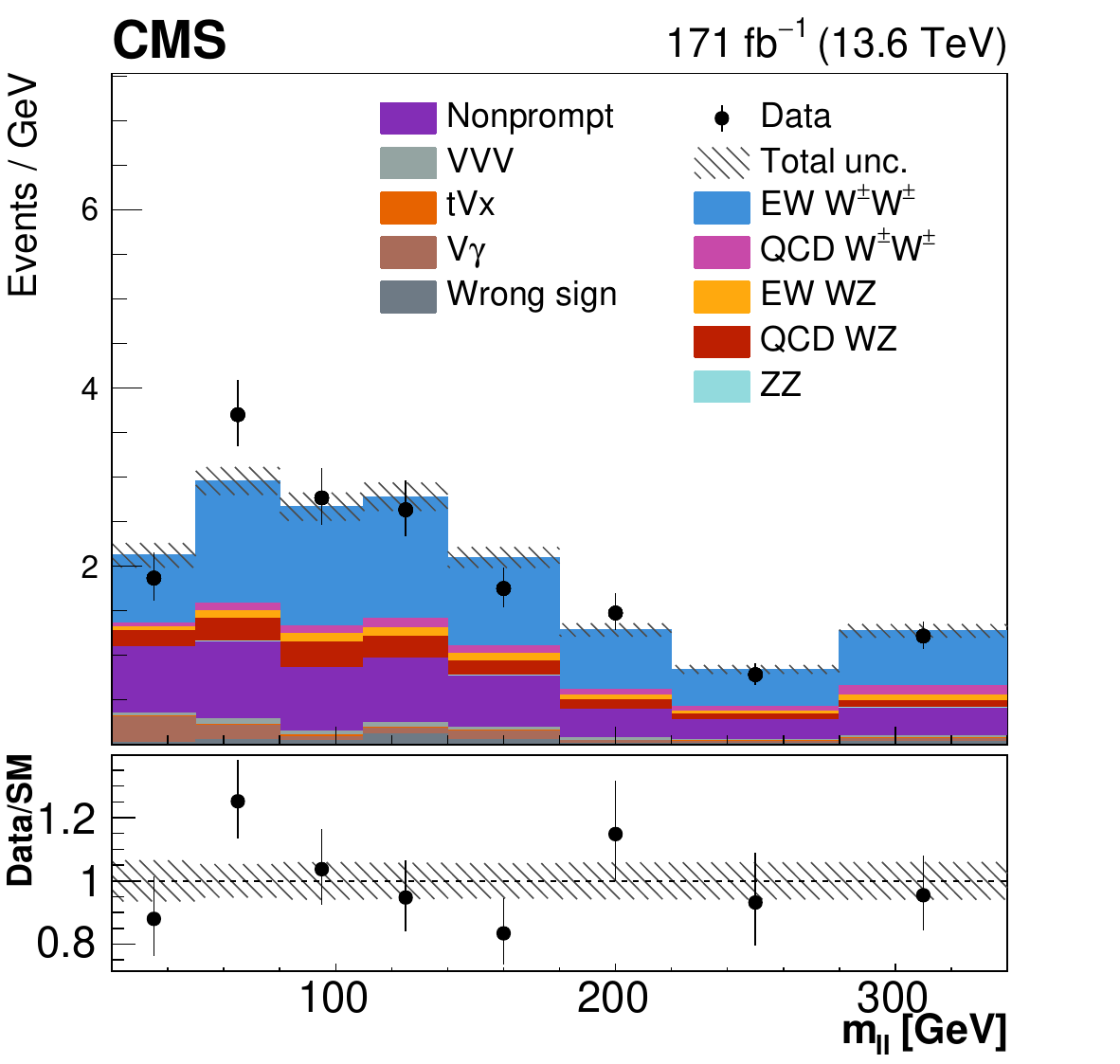}
\includegraphics[width=0.41\textwidth]{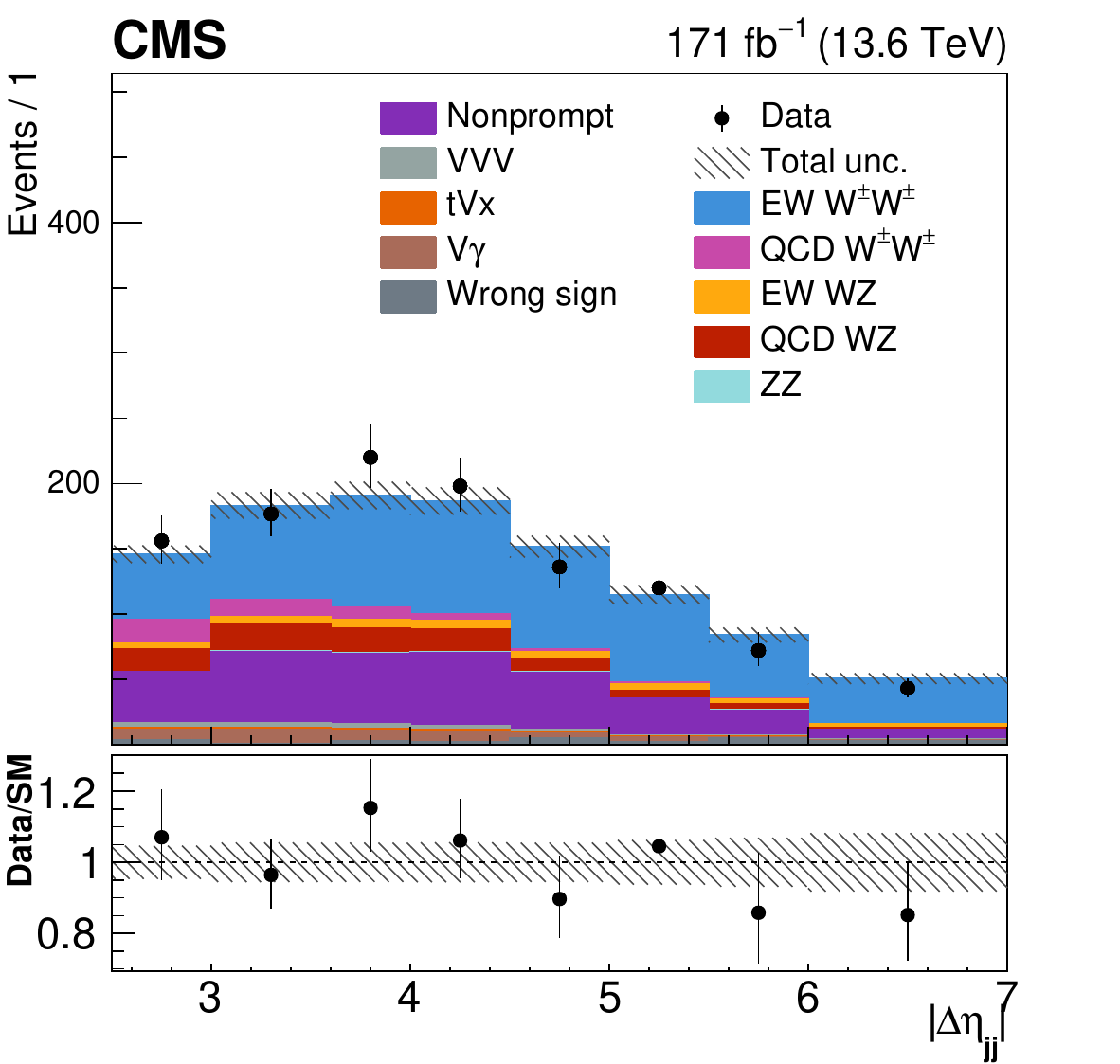}
\includegraphics[width=0.41\textwidth]{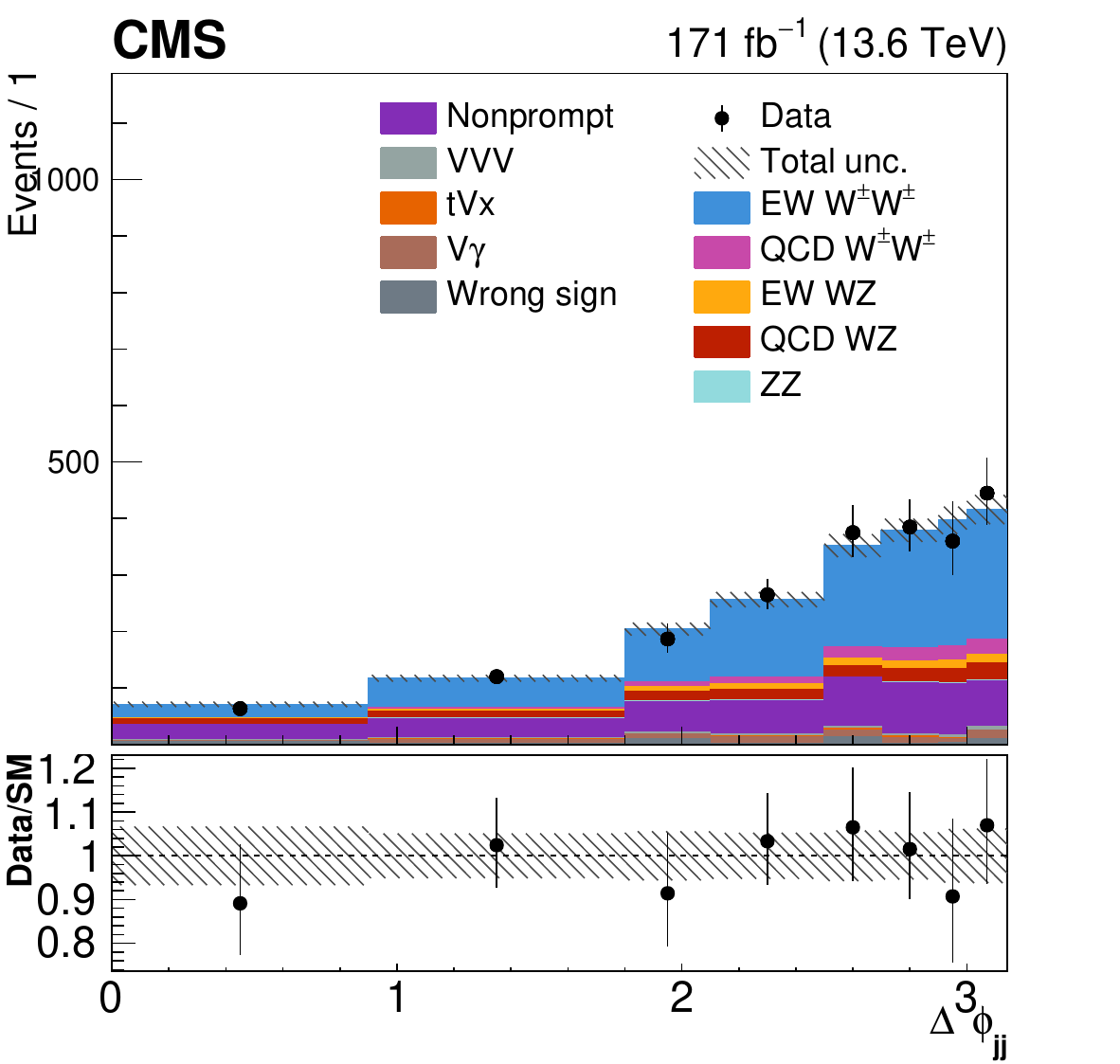}
\includegraphics[width=0.41\textwidth]{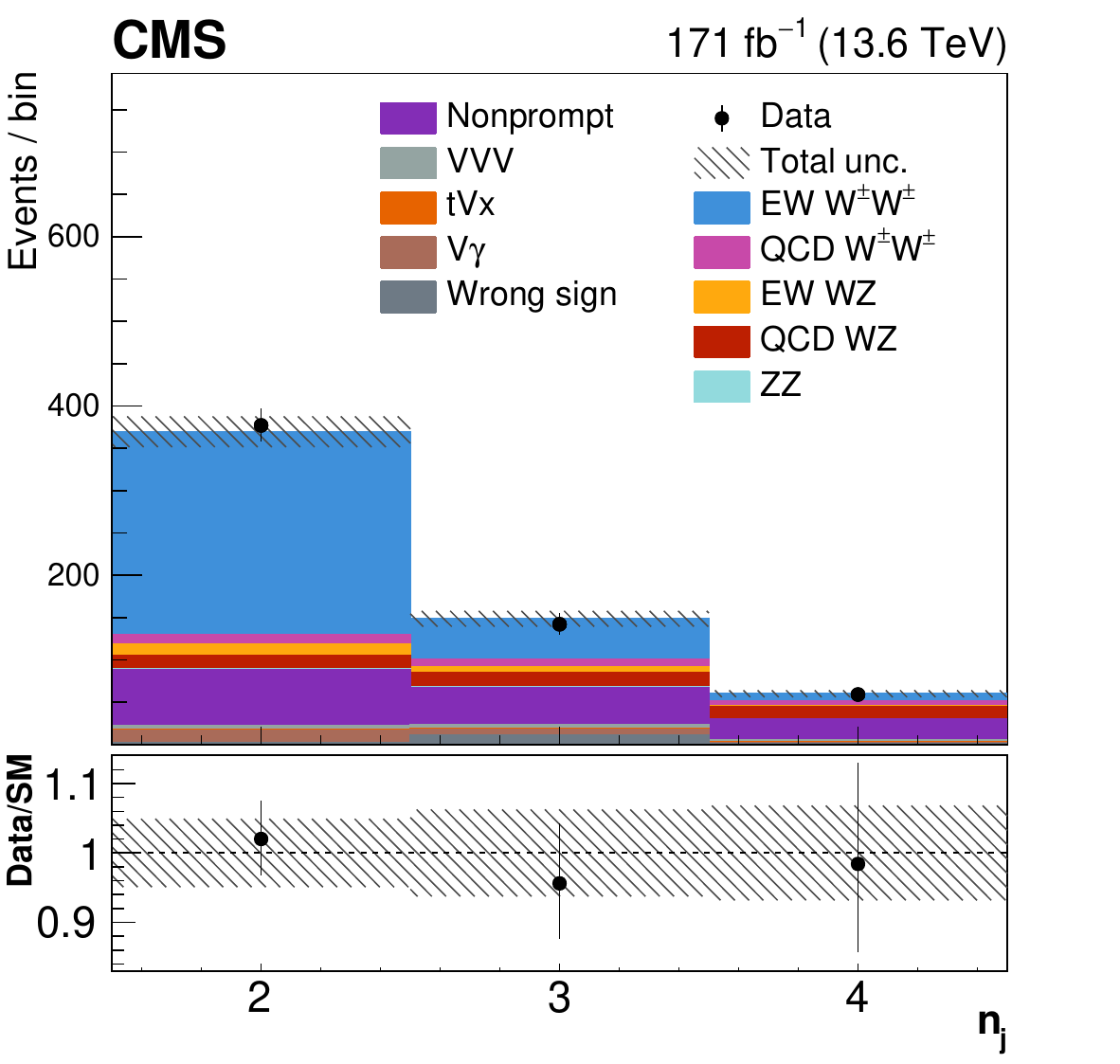}
\caption{Distributions of $\mjj$ (upper left), $\mll$ (upper right), $\detajj$ (center left), $\dphijj$ (center right), and $\NJ$ (lower) in the $\WW$ SR. The predicted yields are shown with their best fit normalizations from the simultaneous fit. Vertical bars on data points represent the statistical uncertainty in the data. The histograms for $\tVx$ backgrounds include the contributions from $\ttbar\PV$ and $\tZq$ processes. The histograms for the wrong-sign background include the contributions from oppositely charged dilepton final states. The overflow is included in the last bin. The bottom panel shows the ratio of the number of events observed in data to that of the total SM prediction. The gray bands represent the total uncertainties in the predicted yields.\label{fig:ssww_signalsel}}
\end{figure*}

\begin{figure*}[htbp]
\centering
\includegraphics[width=0.49\textwidth]{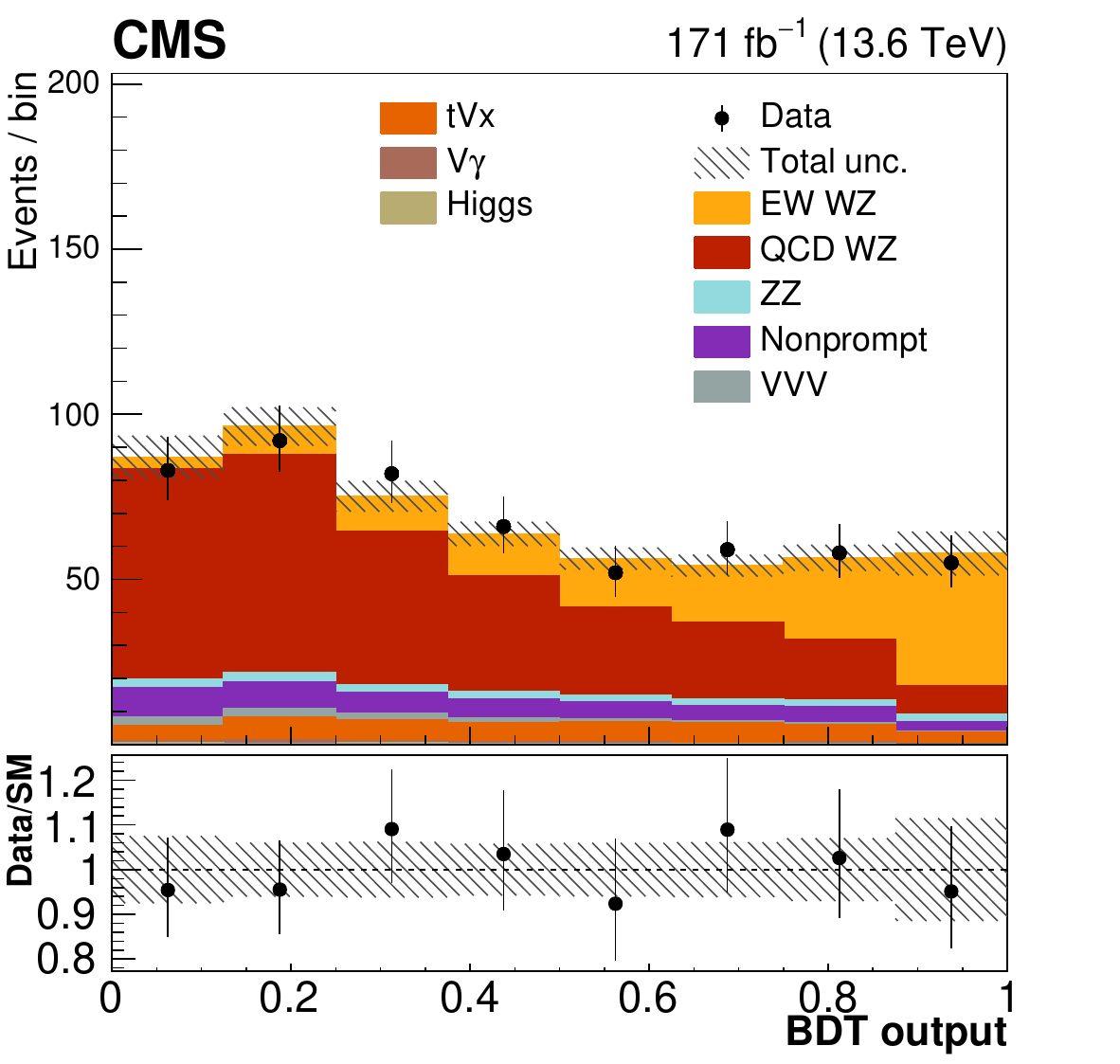}
\includegraphics[width=0.49\textwidth]{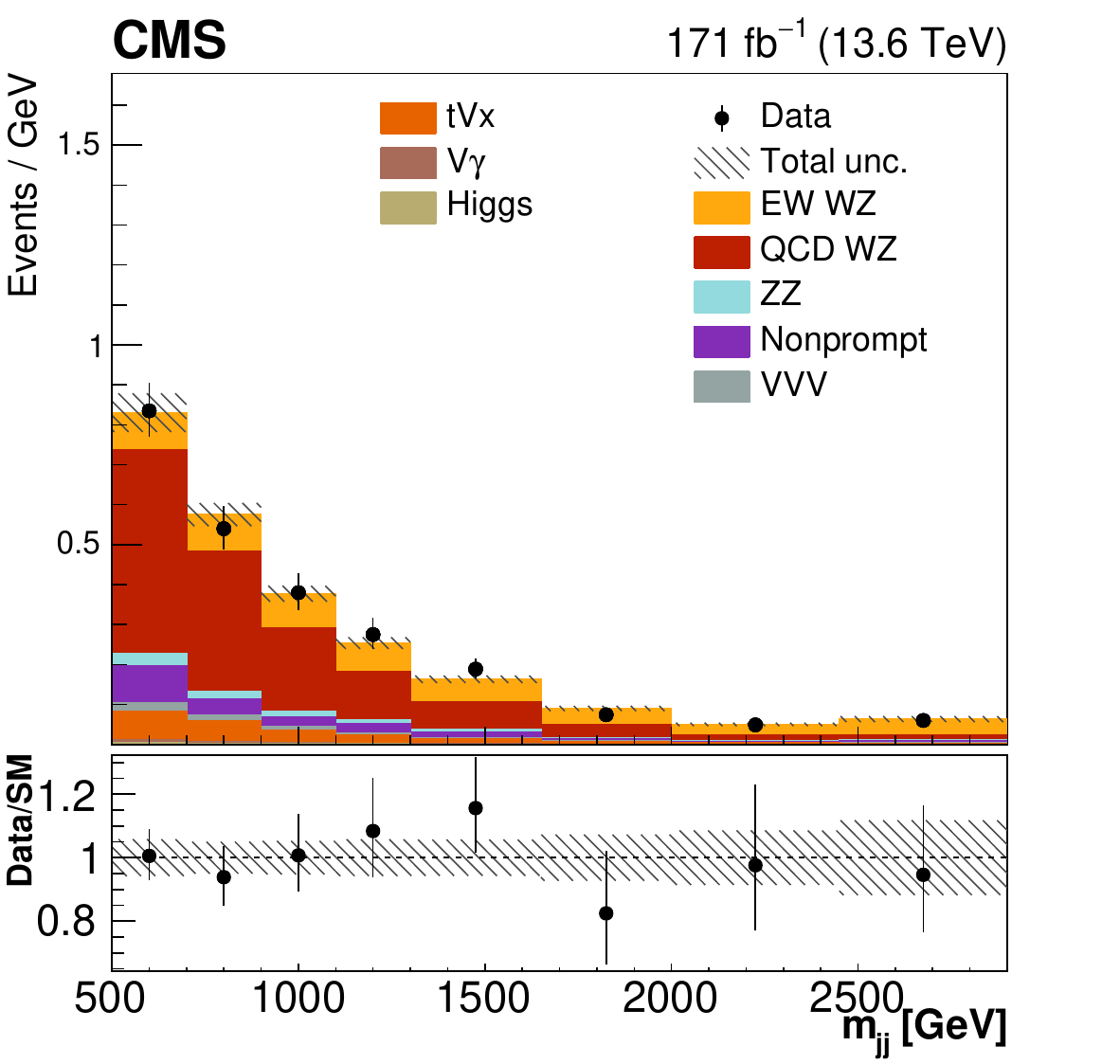}
\caption{Distributions of the BDT score (left) and $\mjj$ (right) in the $\WZ$ SR. The predicted yields are shown 
with their best fit normalizations from the simultaneous fit. Notations are as in Fig.~\ref{fig:ssww_signalsel}.\label{fig:wz_signalsel}}
\end{figure*}

The distributions of $\mjj$, $\mll$, $\detajj$, $\dphijj$, and $\NJ$ in the $\WW$ SR, and the distributions of the BDT score and $\mjj$ in the $\WZ$ SR, are shown in Figs.~\ref{fig:ssww_signalsel} and~\ref{fig:wz_signalsel}, respectively. The data yields, together with the numbers of fitted signal and background events, are given in Table~\ref{tab:yields}. The significances of the EW  $\WW$ and $\WZ$ signals are quantified from the $p$-value using a profile ratio test statistic~\cite{Junk,Read} in an asymptotic approximation~\cite{Cowan:2010js}.  The observed and expected statistical significance from the background-only hypothesis of the EW $\WZ$ signal is about seven standard deviations, while the statistical significance of the EW $\WW$ signal is far above five standard deviations.

\begin{table*}[htbp]
\centering
\topcaption{Expected yields from SM processes and observed data events in the $\WW$ SR, nonprompt-lepton CR, $\WZ$ SR, and $\tZq$ CR. The combination of the statistical and systematic uncertainties is indicated. The expected yields are shown with their best fit normalizations from the simultaneous fit to the data.\label{tab:yields}}
\begin{tabular}{lcccc}
\hline
Process                  &  $\WW$ SR & Nonprompt-lepton CR & $\WZ$ SR & $\tZq$ CR \\
\hline
EW $\WW$       &     282 $\pm$    36 &      98 $\pm$	13 &	 \NA		  &  \NA               \\
QCD $\WW$      &    25.5 $\pm$   3.0 &    17.9 $\pm$   2.0 &	 \NA		  &  \NA               \\
EW $\WZ$       &    20.8 $\pm$   4.5 &     9.7 $\pm$   1.9 &	 131 $\pm$    25 &      0.8 $\pm$   0.2 \\ 
QCD $\WZ$      &    48.6 $\pm$   9.1 &    40.9 $\pm$   6.0 &	 288 $\pm$    49 &     11.4 $\pm$   2.3 \\ 
$\PZ\PZ$     &     1.6 $\pm$   0.3 &     1.3 $\pm$   0.2 &	18.8 $\pm$   3.8 &      0.8 $\pm$   0.1 \\ 
Nonprompt      &     142 $\pm$    39 &     348 $\pm$	47 &	  47 $\pm$    23 &     17.0 $\pm$   8.4 \\ 
$\PV\PV\PV$    &     9.8 $\pm$   2.6 &    20.8 $\pm$   3.9 &	11.3 $\pm$   2.6 &      0.4 $\pm$   0.1 \\
$\tVx$         &     3.7 $\pm$   1.4 &    49.7 $\pm$   8.4 &	  44 $\pm$    13 &      102 $\pm$    16 \\ 
$\PW\PGg$    &      23 $\pm$    10 &    12.5 $\pm$   4.2 &	 4.6 $\pm$   2.3 &      0.2 $\pm$   0.1 \\ 
Wrong-sign     &    13.6 $\pm$   7.1 &    36.5 $\pm$   9.8 & \NA		 & \NA		        \\ 
Higgs boson    & \NA		     & \NA		   &	 2.6 $\pm$   1.3 &      1.4 $\pm$   0.3 \\

Total SM       &     571 $\pm$    24 &     636 $\pm$	25 &	 548 $\pm$    23 &      134 $\pm$    12 \\[\cmsTabSkip]

Data           &    578              &    629              &    547              &    135               \\
\hline
  \end{tabular}
\end{table*}

The fiducial region is defined by a common set of kinematic requirements on the leptons at the generator level, emulating the selection performed at the reconstruction level. The measured distributions, after subtracting the contributions from the background processes, are corrected for detector resolution effects and inefficiencies. The leptons at the generator level are selected at the so-called ``dressed level" by combining the four-momentum of each lepton after final-state photon radiation with that of photons found within a cone of $\Delta R=0.1$ around the lepton. The $\WW$ fiducial region is defined by requiring two same-charge leptons with $\pt>20\GeV$, $\abs{\eta}<2.5$, and $\mll>20\GeV$, and two jets with $\mjj>500\GeV$ and $\detajj>2.5$. The jets at generator level are clustered from stable particles, excluding neutrinos, using the anti-$\kt$ clustering algorithm with a distance parameter of 0.4, and are required to have $\pt>50\GeV$ and $\abs{\eta}<4.7$. The jets within $\Delta R<0.4$ of the selected charged leptons are not included. The $\WZ$ fiducial region is defined by requiring three leptons with $\pt>20\GeV$ and $\abs{\eta}<2.5$, among which there is a pair of opposite-charge same-flavor lepton pair with $\abs{\mll-\mZ}<15\GeV$, and two jets with $\mjj>500\GeV$ and $\detajj>2.5$. The \MGvATNLO generator is used to extrapolate from the reconstruction level to the fiducial phase space. Electrons and muons produced in the decay of a $\Pgt$ lepton are not included in the definition of the fiducial region. All fiducial requirements are listed in Table~\ref{tab:fiducial}. Nonfiducial events, $\ie$, events selected at the reconstructed level that do not satisfy the fiducial requirements, are included as background processes in the simultaneous fit.

\begin{table}[htbp]
  \centering
  \topcaption{
    Definition of the fiducial regions. \label{tab:fiducial}
  }
  \begin{tabular}{lcc}
    \hline
    Observable            & $\WW$			  & $\WZ$ \\
    \hline
    Lepton origin         & Direct decay of a \PW boson & Direct decay of a \PW or \PZ boson \\
    Lepton definition     & \multicolumn{2}{c}{Dressed leptons} \\
    Leptons selection     & 2 leptons, $\pt>25/20\GeV$ & 3 leptons, $\pt>25/20/20\GeV$ \\
    $\abs{\eta_{\Pell}}$   & \multicolumn{2}{c}{$<$2.5} \\
    $\mll$                & $>$20\GeV & \NA \\
    $\abs{\mll-\mZ}$      & \NA & $<$15\GeV \\
    $\pt^\jet$            & \multicolumn{2}{c}{$>$50\GeV} \\
    $\abs{\eta^\jet}$     & \multicolumn{2}{c}{$<$4.7} \\
    $\mjj$                & \multicolumn{2}{c}{$>$500\GeV} \\
    $\detajj$             & \multicolumn{2}{c}{$>$2.5} \\
    $\Delta R(\jet,\Pell)$ & \multicolumn{2}{c}{$>$0.4} \\
    \hline
  \end{tabular}
\end{table}

In addition to the fiducial cross sections for the EW production, inclusive fiducial cross sections are also reported for the combined EW+QCD production. Here, "EW+QCD" denotes the total diboson-plus-two-jets production in the fiducial region, obtained as the sum of the EW and QCD-induced contributions, with the EW-QCD interference assigned to the QCD component as described in Section~\ref{sec:samples}. Inclusive cross section measurements for the EW $\WW$, EW+QCD $\WW$, EW $\WZ$, and EW+QCD $\WZ$ processes, together with their theoretical predictions with \MGvATNLO and \SHERPA 2.2.15~\cite{Sherpa:2019gpd} at LO, are given in Table~\ref{tab:default_fid_inc}. The \MGvATNLO predictions including the $\mathcal{O}(\alpS\alpha^6)$ and $\mathcal{O}(\alpha^7)$ corrections in the EW $\WW$ and $\WZ$ processes are also shown. Both the differential cross sections from \MGvATNLO and \SHERPA and the data measurements agree with each other within the theoretical and experimental uncertainties, although \SHERPA tends to predict lower central values. In comparison with the results from Ref.~\cite{CMS:2020gfh} at $\sqrt{s}=13\TeV$, the uncertainties are about 10--15\% lower.

\begin{figure*}[htbp]
\centering
\includegraphics[width=0.41\textwidth]{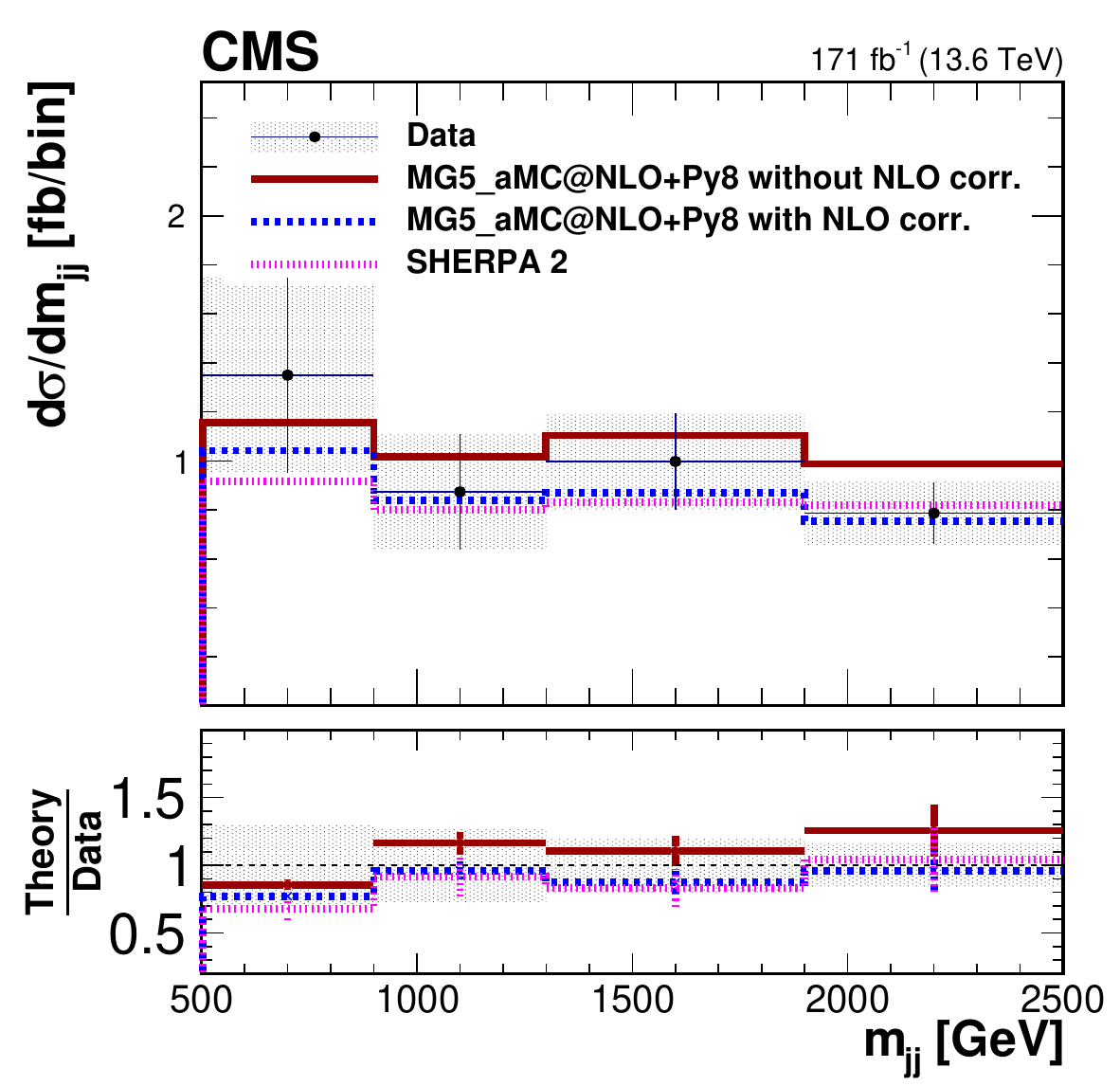}
\includegraphics[width=0.41\textwidth]{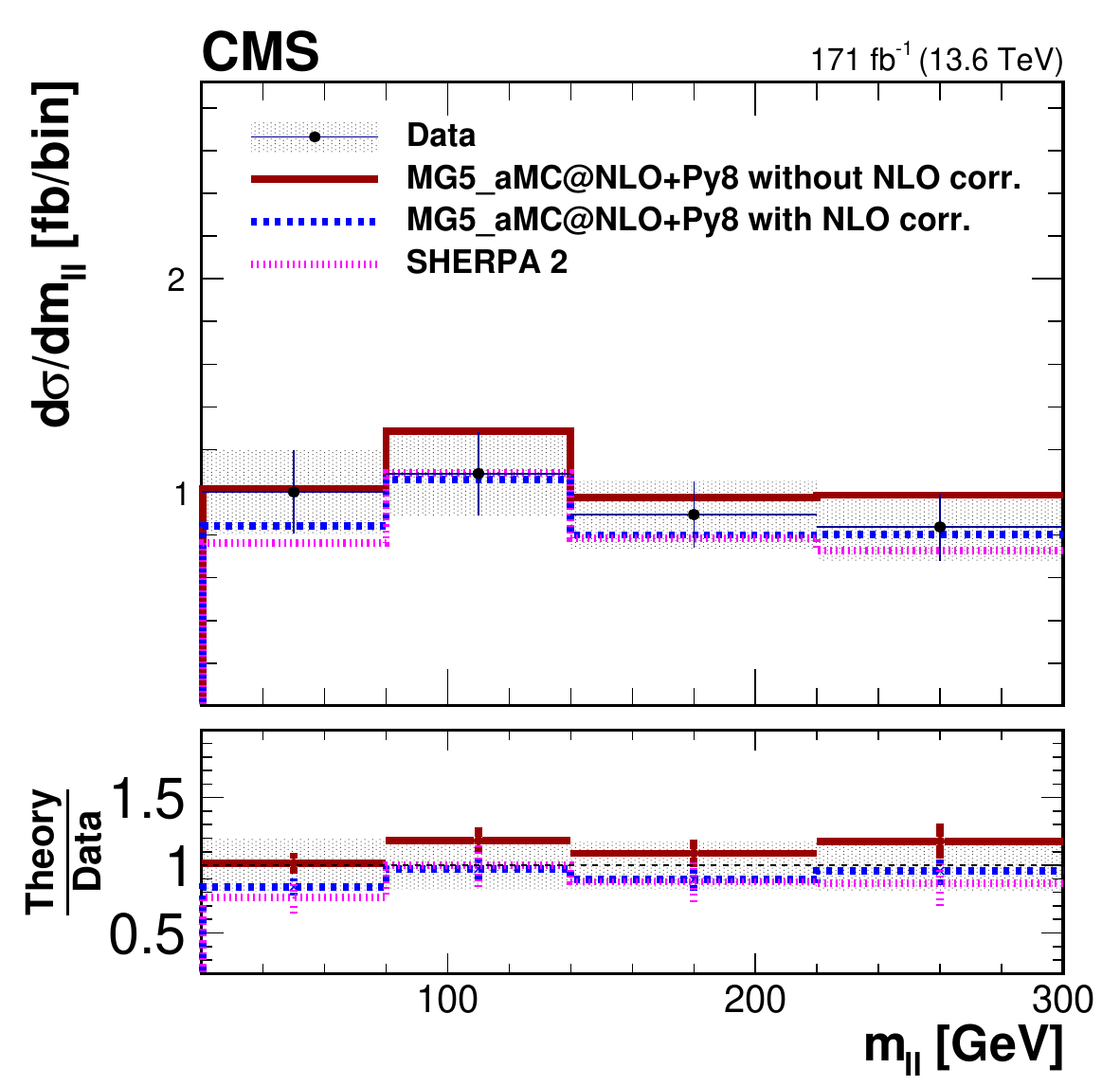}
\includegraphics[width=0.41\textwidth]{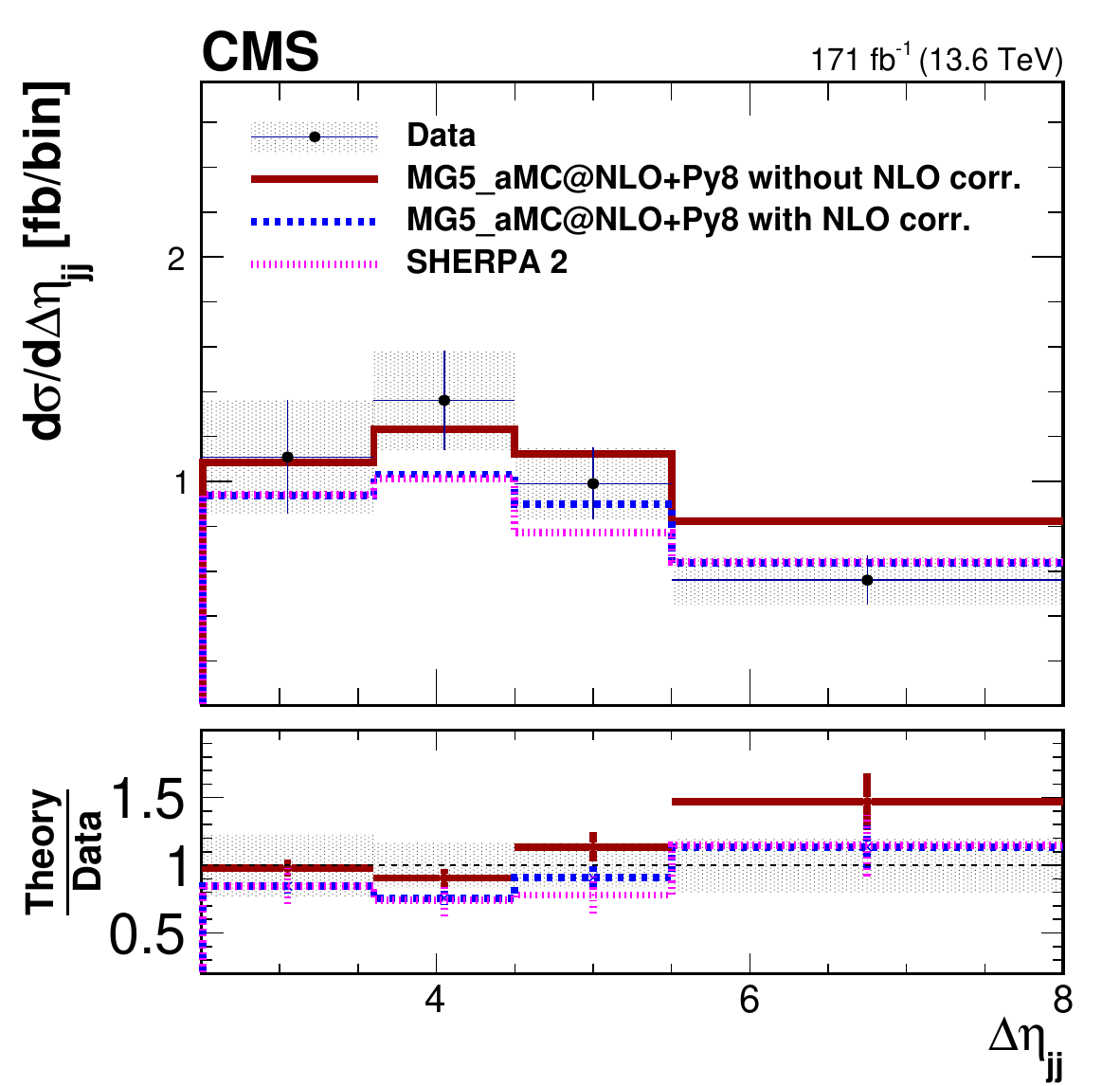}
\includegraphics[width=0.41\textwidth]{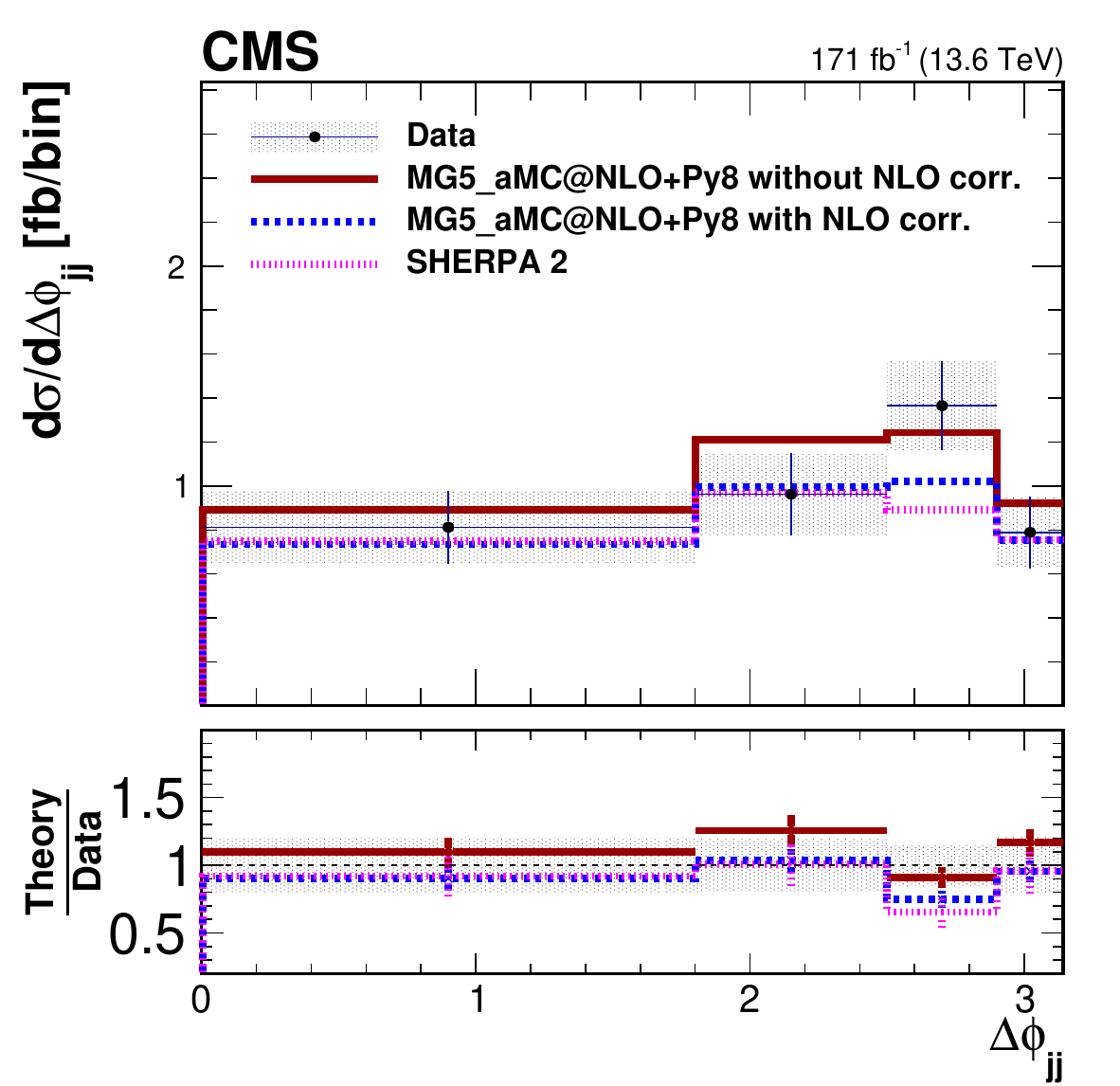}
\includegraphics[width=0.41\textwidth]{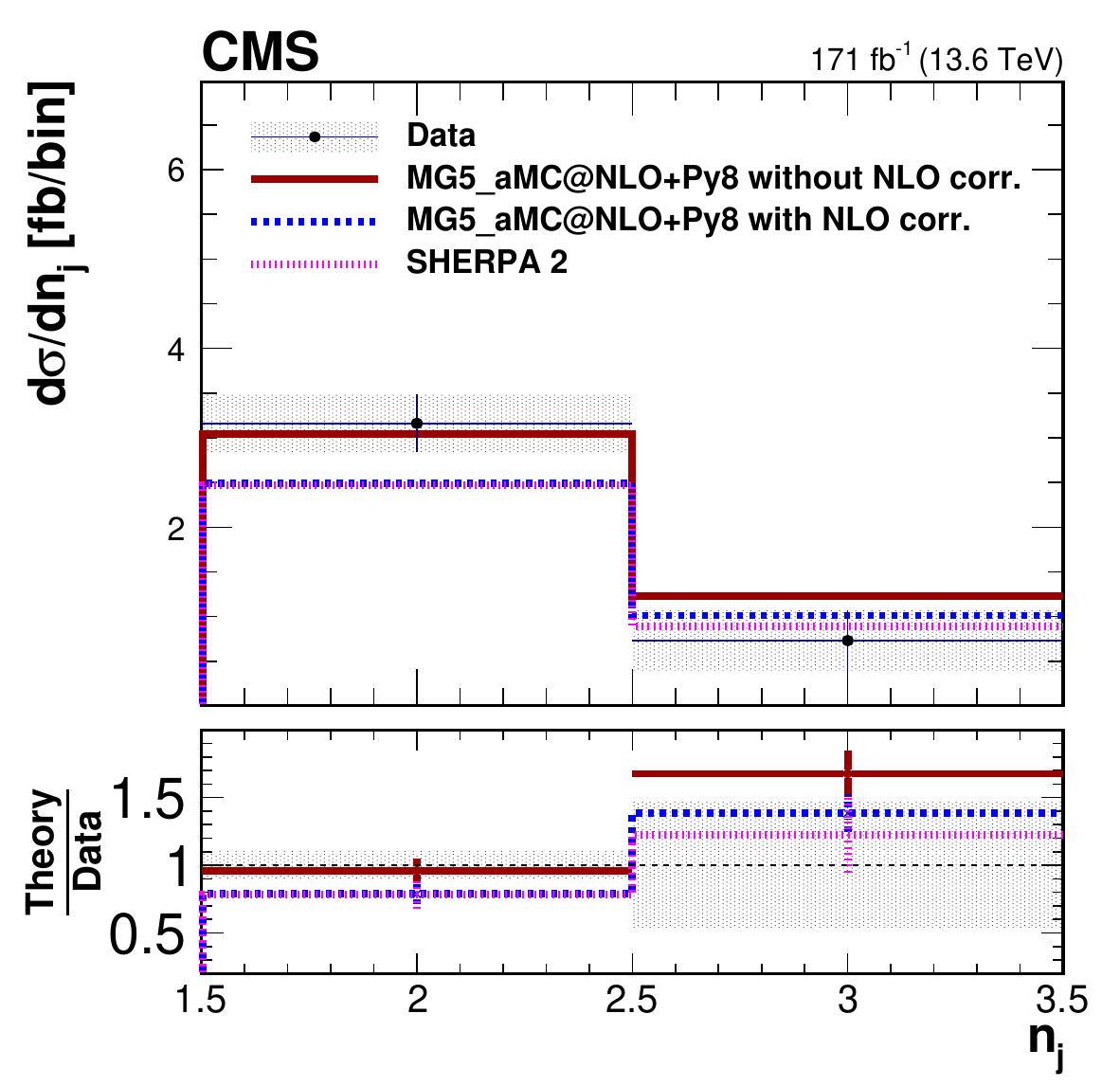}
\caption{Measured and predicted absolute EW $\WW$ cross sections in bins of $\mjj$ (upper left), $\mll$ (upper right), $\detajj$ (center left), $\dphijj$ (center right), and $\NJ$ (lower). The ratios of the predictions to the data are also shown. The measurements are compared with the predictions from the \MGvATNLO and \SHERPA generators. The shaded bands and uncertainty bars around the data points correspond to the total measurement uncertainty. The uncertainty bars around the predictions correspond to the combined statistical, PDF, and scale uncertainties. Predictions without applying the $\mathcal{O}(\alpS\alpha^6)$ and $\mathcal{O}(\alpha^7)$ corrections to the \MGvATNLO LO cross sections, as described in the text, are also shown.\label{fig:unf_ww_normalized0}}
\end{figure*}

\begin{figure*}[htbp]
\centering
\includegraphics[width=0.41\textwidth]{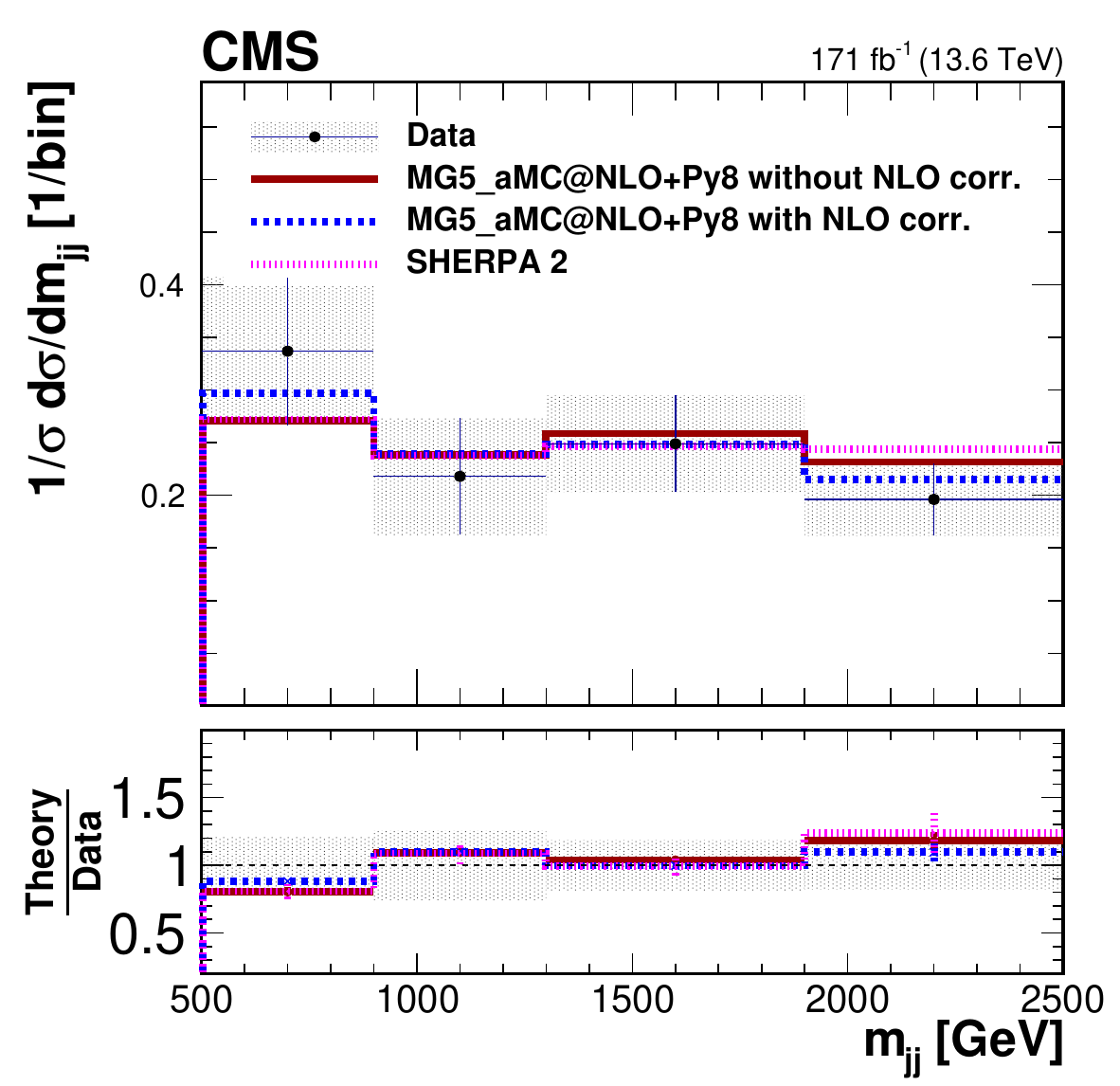}
\includegraphics[width=0.41\textwidth]{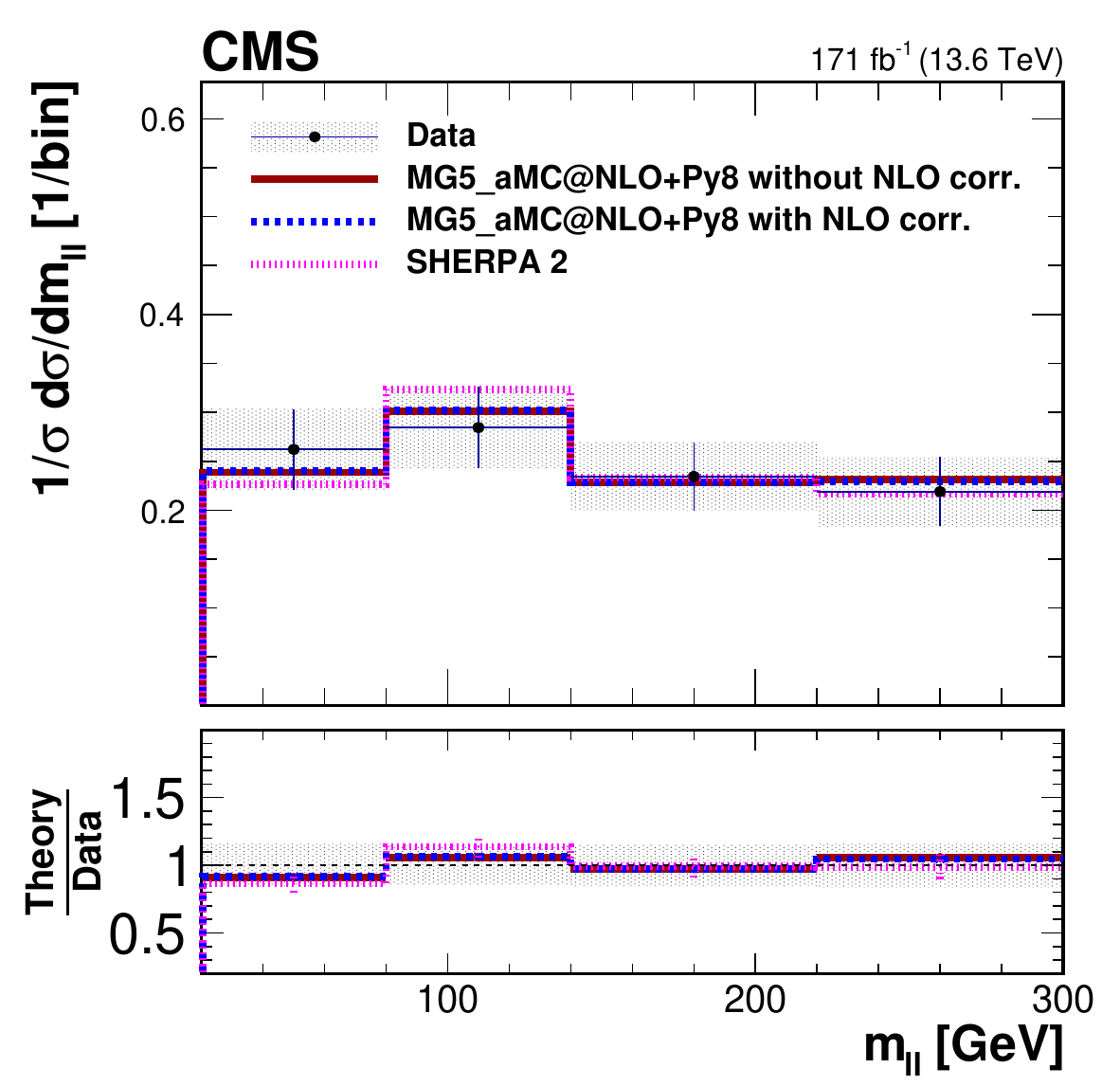}
\includegraphics[width=0.41\textwidth]{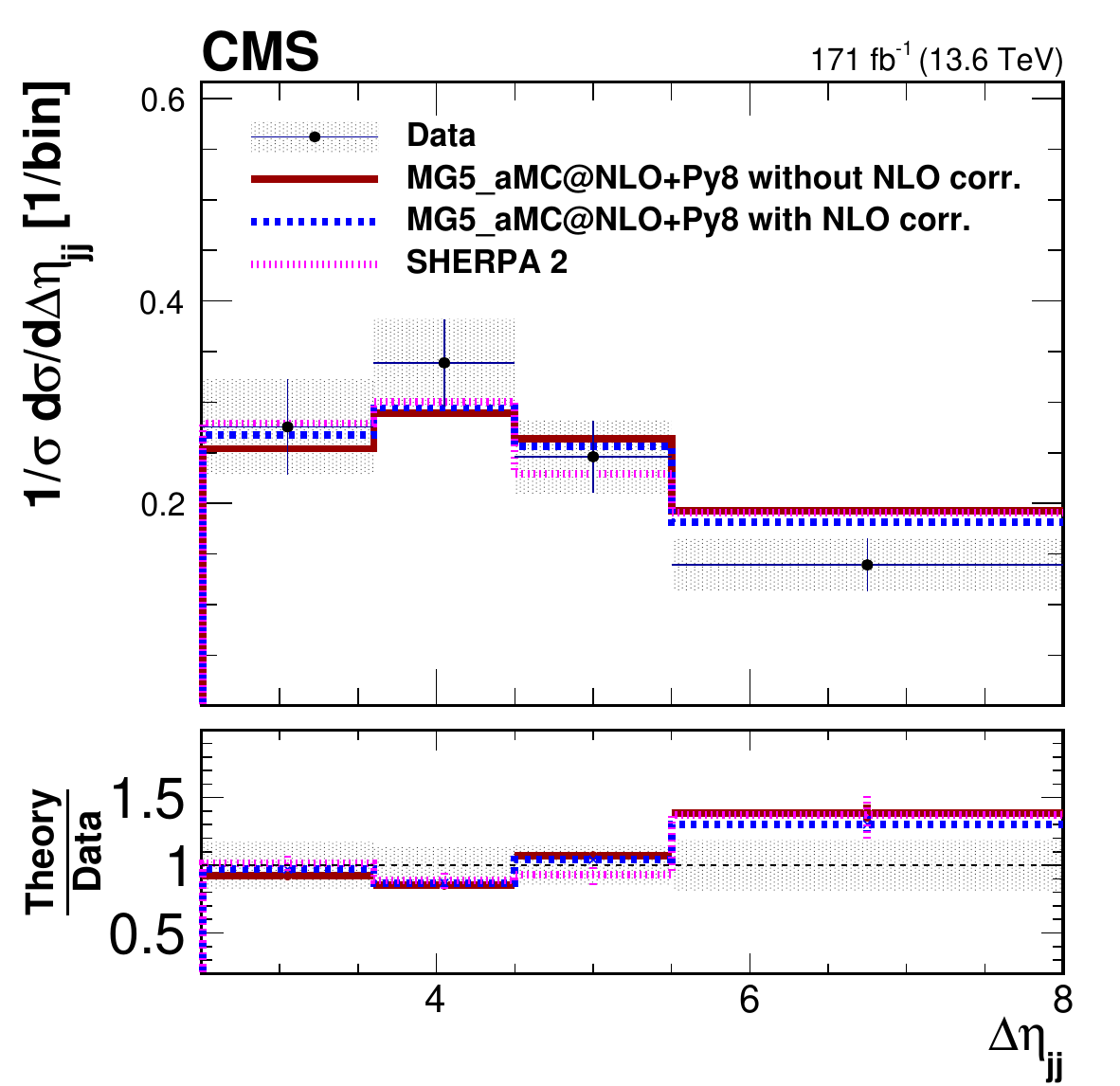}
\includegraphics[width=0.41\textwidth]{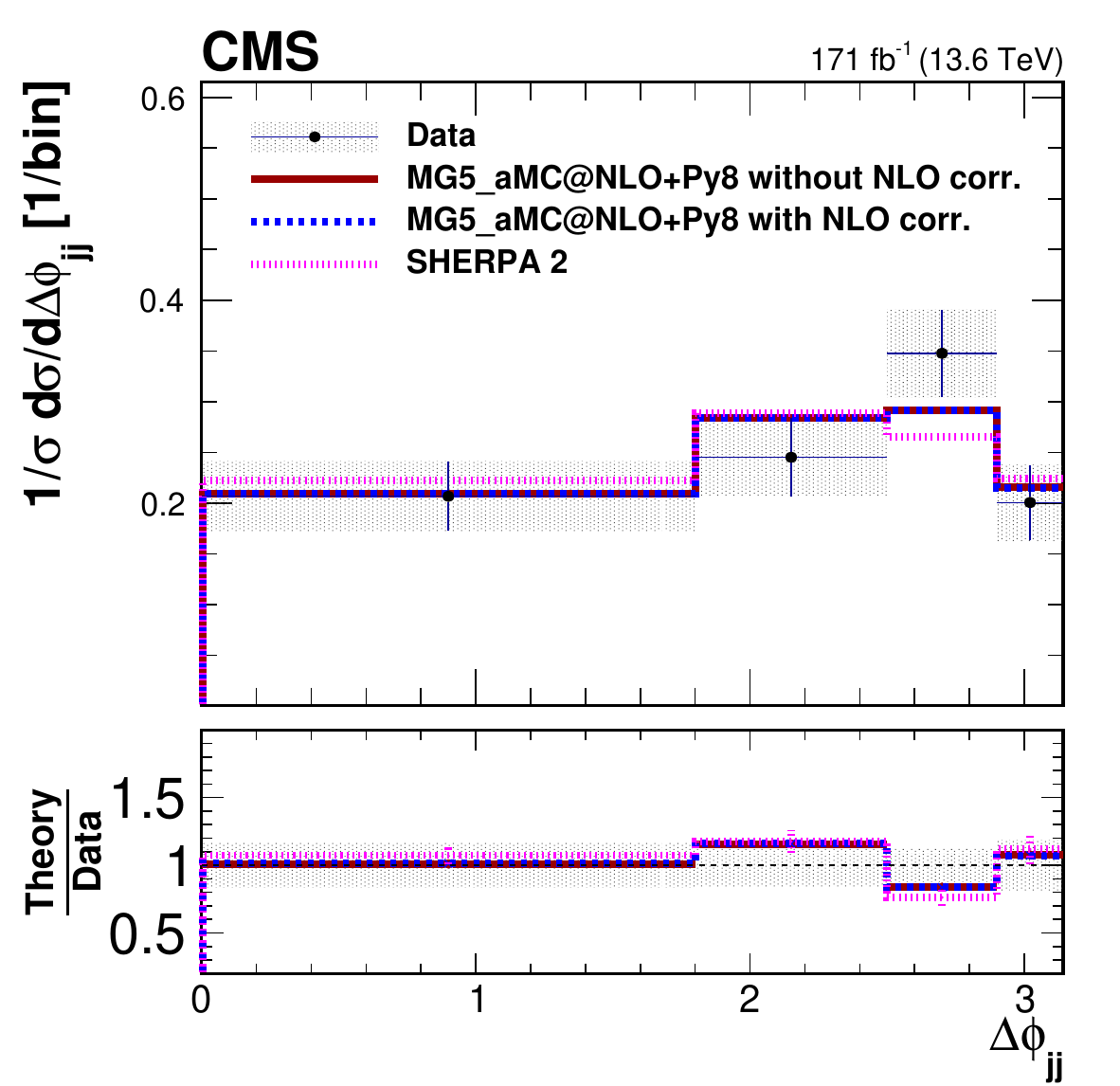}
\includegraphics[width=0.41\textwidth]{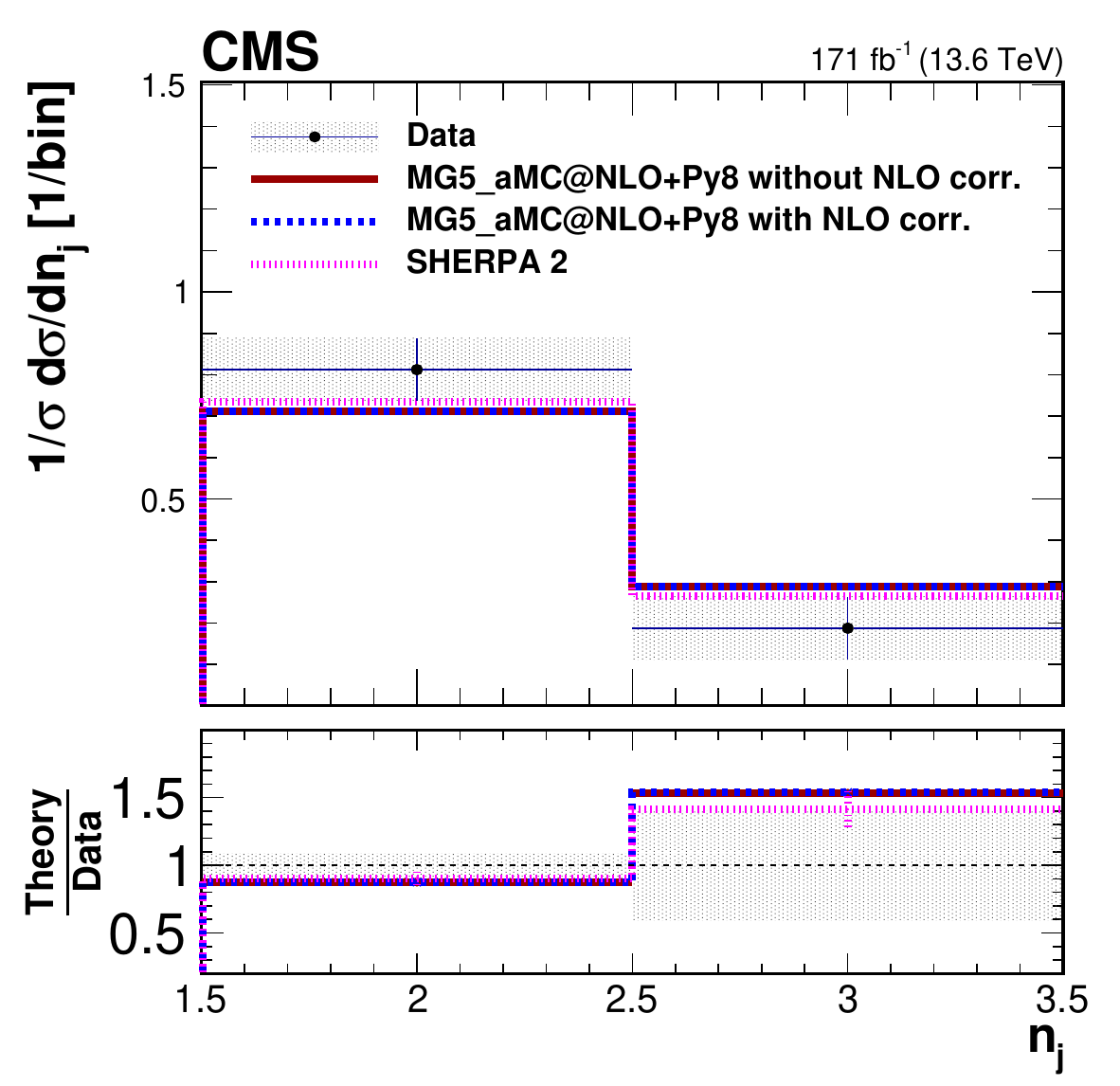}
\caption{Measured and predicted normalized EW $\WW$ cross sections in bins of $\mjj$ (upper left), $\mll$ (upper right), $\detajj$ (center left), $\dphijj$ (center right), and $\NJ$ (lower). The ratios of the predictions to the data are also shown. The measurements are compared with the predictions from the \MGvATNLO and \SHERPA generators. The shaded bands and uncertainty bars around the data points correspond to the total measurement uncertainty. The uncertainty bars around the predictions correspond to the combined statistical, PDF, and scale uncertainties. Predictions without applying the $\mathcal{O}(\alpS\alpha^6)$ and $\mathcal{O}(\alpha^7)$ corrections to the \MGvATNLO LO cross sections, as described in the text, are also shown.\label{fig:unf_ww_normalized1}}
\end{figure*}

\begin{figure*}[htb!]
\centering
\includegraphics[width=0.41\textwidth]{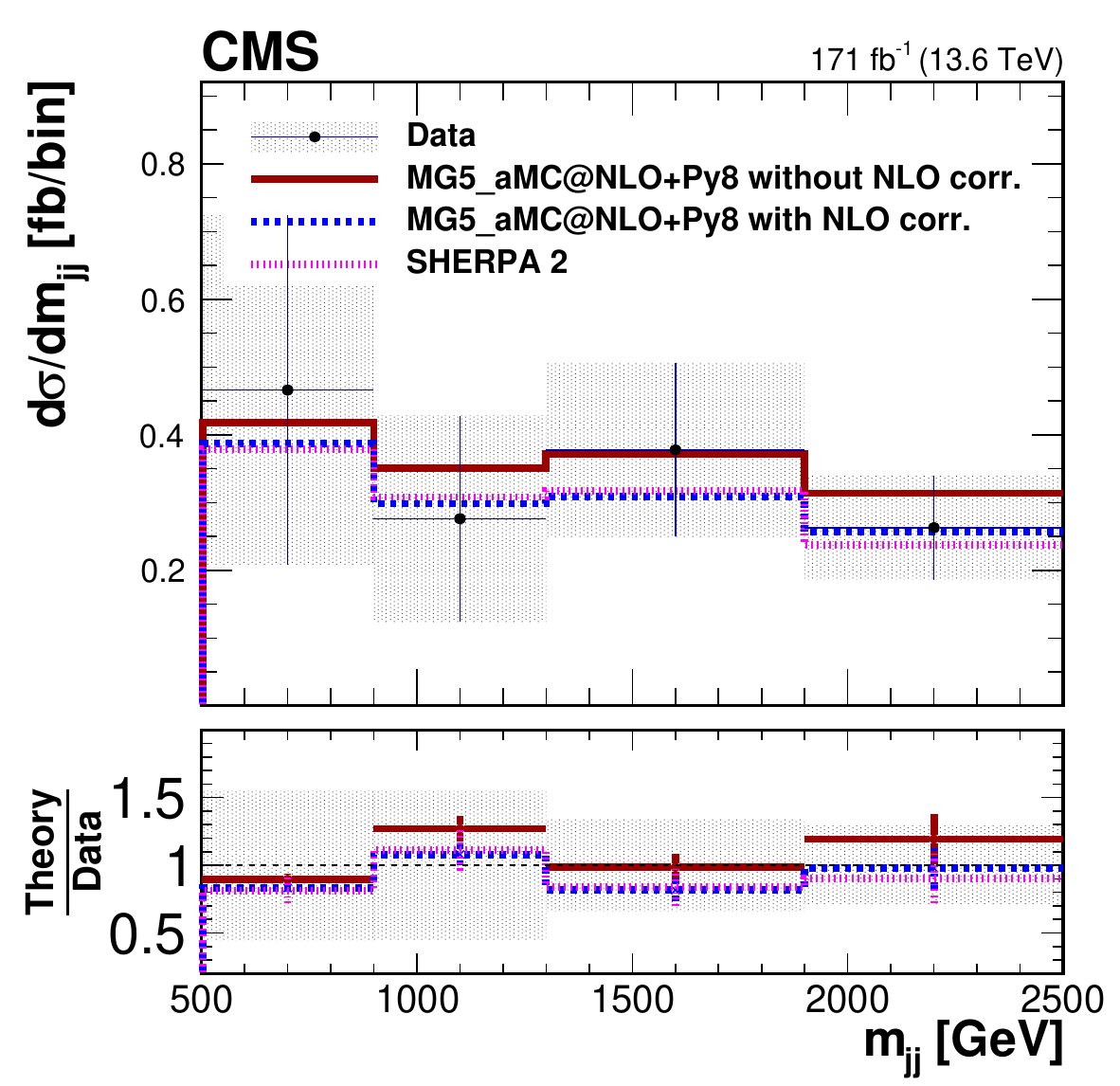}
\includegraphics[width=0.41\textwidth]{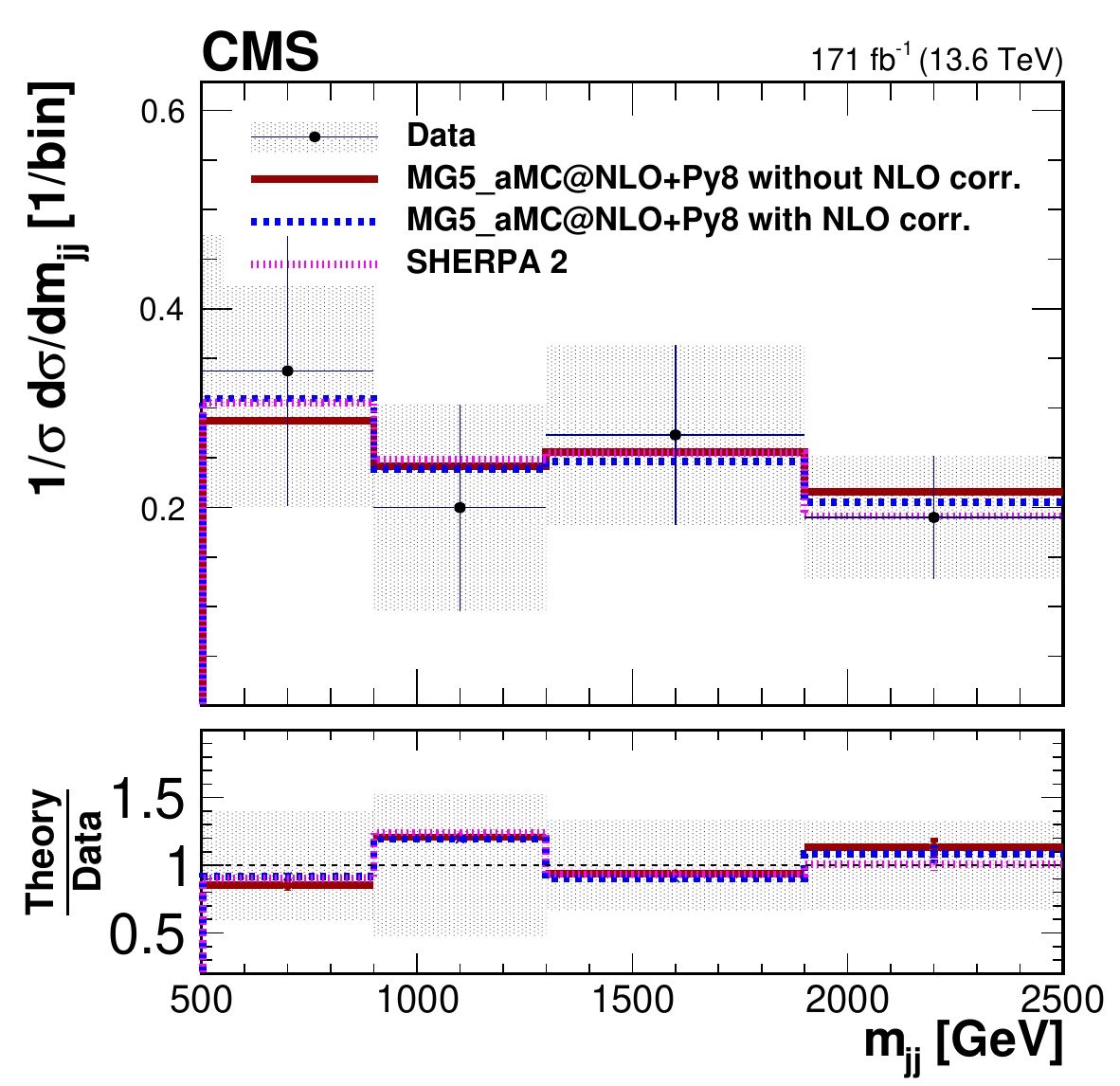}
\caption{Measured and predicted absolute (left) and normalized (right) EW $\WZ$ cross sections in bins of $\mjj$. The ratios of the predictions to the data are also shown. The measurements are compared with the predictions from the \MGvATNLO and \SHERPA generators. The shaded bands and uncertainty bars around the data points correspond to the total measurement uncertainty. The uncertainty bars around the predictions correspond to the combined statistical, PDF, and scale uncertainties. Predictions without applying the $\mathcal{O}(\alpS\alpha^6)$ and $\mathcal{O}(\alpha^7)$ corrections to the \MGvATNLO LO cross sections, as described in the text, are also shown.\label{fig:unf_wz_normalized}}
\end{figure*}

To perform absolute differential production cross section measurements, signal templates are built from different bins of differential-basis
observable values predicted by the event generator. Each signal template is considered as a separate process in the simultaneous binned maximum-likelihood fit. The signal extraction at reconstruction level and the unfolding into the generator level bins are performed in a single step. The bin migration effects due to the detector resolution are negligible. In this paper, measurements of variables sensitive to EW production are reported. The measured EW $\WW$ differential cross sections in bins of $\mjj$, $\mll$, $\NJ$, $\detajj$, and $\dphijj$ are shown in Fig.~\ref{fig:unf_ww_normalized0}, while the corresponding normalized cross sections are shown in Fig.~\ref{fig:unf_ww_normalized1}. The $\NJ$, $\detajj$, and $\dphijj$ differential cross section measurements are performed by replacing the $\mll$ variable by the corresponding quantity in the $\WW$ SR in the simultaneous fit. The measured EW $\WZ$ differential and normalized cross sections in bins of $\mjj$ are shown in Fig.~\ref{fig:unf_wz_normalized}. The normalized differential cross sections agree well between predictions and observed data.

\begin{table*}[htbp]
\centering
\topcaption{
The measured inclusive cross sections multiplied by the branching fractions ($\sigma \, \mathcal{B}$) of the $\PW (\PZ)$ boson to $\Pell\PGn(\Pell\Pell)$ for the EW $\WW$, EW+QCD $\WW$, EW $\WZ$, and EW+QCD $\WZ$ processes, and the theoretical predictions with \MGvATNLO and \SHERPA at LO. Here, EW+QCD refers to the sum of the electroweak and QCD-induced production in the fiducial region, with the EW-QCD interference included in the QCD component. The theoretical uncertainties include statistical, PDF, and scale uncertainties. Predictions with applying the $\mathcal{O}(\alpS\alpha^6)$ and $\mathcal{O}(\alpha^7)$ corrections to the \MGvATNLO LO cross sections, as described in the text, are also shown. The predictions of the QCD $\WW$ and $\WZ$ processes do not include additional corrections. All reported values are in \unit{fb}.\label{tab:default_fid_inc}}
\cmsTable{
\begin{tabular}{ccccc}
\hline
\multicolumn{2}{c}{}                                                            & \multicolumn{3}{c}{Predictions (\unit{fb})} \\
\multirow{2}{*}{Process} & \multirow{2}{*}{$\sigma \, \mathcal{B}$ (\unit{fb})} & \MGvATNLO & \MGvATNLO & \SHERPA \\
			 &                                                      & without NLO corrections & with NLO corrections & without NLO corrections \\
\hline
\multirow{2}{*}{EW $\WW$}     &  $3.81 \pm 0.38$            &  \multirow{2}{*}{$4.27 \pm 0.38$} &  \multirow{2}{*}{$3.51 \pm 0.31$} &  \multirow{2}{*}{$3.37 \pm 0.50$} \\
                              &  $0.33\stat \pm 0.18\syst$  &					&				    &					\\
\multirow{2}{*}{EW+QCD $\WW$} &  $4.32\pm 0.40$             &  \multirow{2}{*}{$4.75 \pm 0.52$} &  \multirow{2}{*}{$3.99 \pm 0.44$} &  \NA                              \\
                              &  $0.36\stat \pm 0.18\syst$  &					&				    &					\\
\multirow{2}{*}{EW $\WZ$}     &  $1.43\pm 0.26$ 	    &  \multirow{2}{*}{$1.45 \pm 0.13$} &  \multirow{2}{*}{$1.25 \pm 0.11$} &  \multirow{2}{*}{$1.24 \pm 0.19$} \\
                              &  $0.23\stat \pm 0.12\syst$  &					&				    &					\\
\multirow{2}{*}{EW+QCD $\WZ$} &  $4.75\pm 0.35$             &  \multirow{2}{*}{$4.59 \pm 1.07$} &  \multirow{2}{*}{$4.39 \pm 1.05$} &  \NA                              \\
                              &  $0.27\stat \pm 0.22\syst$  &					&				    &					\\
\hline
\end{tabular}
}
\end{table*}

\clearpage

\section{Summary}
\label{sec:summary}

The production cross sections of $\WZ$ and same-sign $\WW$ boson pairs in association with two jets are measured for the first time in proton-proton collisions at a center-of-mass energy of 13.6\TeV. The data sample corresponds to an integrated luminosity of \lumiTotal\fbinv, collected with the CMS detector during 2022--2024. The measurements are performed in the leptonic decay modes $\PWpm\PZ \to \Pellpm\PGn\Pellpmpr\Pellmppr$ and $\WW \to \Pellpm\PGn\Pellpmpr\PGn$, where $\Pell, \Pellpr = \Pe $ or $ \PGm$. Electroweak $\WW$ and $\WZ$ fiducial cross sections are measured inclusively and differentially as a function of several kinematic observables and agree with the standard model predictions.

\begin{acknowledgments}
We congratulate our colleagues in the CERN accelerator departments for the excellent performance of the LHC and thank the technical and administrative staffs at CERN and at other CMS institutes for their contributions to the success of the CMS effort. In addition, we gratefully acknowledge the computing centers and personnel of the Worldwide LHC Computing Grid and other centers for delivering so effectively the computing infrastructure essential to our analyses. Finally, we acknowledge the enduring support for the construction and operation of the LHC, the CMS detector, and the supporting computing infrastructure provided by the following funding agencies: SC (Armenia), BMBWF and FWF (Austria); FNRS and FWO (Belgium); CNPq, CAPES, FAPERJ, FAPERGS, and FAPESP (Brazil); MES and BNSF (Bulgaria); CERN; CAS, MoST, and NSFC (China); MINCIENCIAS (Colombia); MSES and CSF (Croatia); RIF (Cyprus); SENESCYT (Ecuador); ERC PRG and PSG, TARISTU24-TK10 and MoER TK202 (Estonia); Academy of Finland, MEC, and HIP (Finland); CEA and CNRS/IN2P3 (France); SRNSF (Georgia); BMFTR, DFG, and HGF (Germany); GSRI (Greece); MATE and NKFIH (Hungary); DAE and DST (India); IPM (Iran); SFI (Ireland); INFN (Italy); MSIT and NRF (Republic of Korea); MES (Latvia); LMTLT (Lithuania); MOE and UM (Malaysia); BUAP, CINVESTAV, CONACYT, LNS, SEP, and UASLP-FAI (Mexico); MOS (Montenegro); MBIE (New Zealand); PAEC (Pakistan); MSHE, NSC, and NAWA (Poland); FCT (Portugal);  MESTD (Serbia); MICIU/AEI and PCTI (Spain); MOSTR (Sri Lanka); Swiss Funding Agencies (Switzerland); MST (Taipei); MHESI (Thailand); TUBITAK and TENMAK (T\"{u}rkiye); NASU (Ukraine); STFC (United Kingdom); DOE and NSF (USA).

\hyphenation{Rachada-pisek} Individuals have received support from the Marie-Curie program and the European Research Council and Horizon 2020 Grant, contract Nos.\ 675440, 724704, 752730, 758316, 765710, 824093, 101115353, 101002207, 101001205, and COST Action CA16108 (European Union); the Leventis Foundation; the Alfred P.\ Sloan Foundation; the Alexander von Humboldt Foundation; the Science Committee, project no. 22rl-037 (Armenia); the Fonds pour la Formation \`a la Recherche dans l'Industrie et dans l'Agriculture (FRIA) and Fonds voor Wetenschappelijk Onderzoek contract No. 1228724N (Belgium); the Beijing Municipal Science \& Technology Commission, No. Z191100007219010, the Fundamental Research Funds for the Central Universities, the Ministry of Science and Technology of China under Grant No. 2023YFA1605804, the Natural Science Foundation of China under Grant No. 12535004, and USTC Research Funds of the Double First-Class Initiative No.\ YD2030002017 (China); the Ministry of Education, Youth and Sports (MEYS) of the Czech Republic; the Shota Rustaveli National Science Foundation (Georgia); the Deutsche Forschungsgemeinschaft (DFG), among others, under Germany's Excellence Strategy -- EXC 2121 ``Quantum Universe" -- 390833306, and under project number 400140256 - GRK2497; the Hellenic Foundation for Research and Innovation (HFRI), Project Number 2288 (Greece); the Hungarian Academy of Sciences, the New National Excellence Program - \'UNKP, the NKFIH research grants K 131991, K 138136, K 143460, K 143477, K 147557, K 146913, K 146914, K 147048, TKP2021-NKTA-64, and 2025-1.1.5-NEMZ\_KI-2025-00004, and MATE KKP and KKPCs Research Excellence and Flagship Research Groups grants (Hungary); the Council of Science and Industrial Research, India; ICSC -- National Research Center for High Performance Computing, Big Data and Quantum Computing, FAIR -- Future Artificial Intelligence Research, and CUP I53D23001070006 (Mission 4 Component 1), funded by the NextGenerationEU program, the Italian Ministry of University and Research (MUR) under Bando PRIN 2022 -- CUP I53C24002390006, PRIN PRIMULA 2022RBYK7T (Italy); the Latvian Council of Science; the Ministry of Science and Higher Education, project no. 2022/WK/14, and the National Science Center, contracts Opus 2021/41/B/ST2/01369, 2021/43/B/ST2/01552, 2023/49/B/ST2/03273, and the NAWA contract BPN/PPO/2021/1/00011 (Poland); the Funda\c{c}\~ao para a Ci\^encia e a Tecnologia (Portugal); the National Priorities Research Program by Qatar National Research Fund; MICIU/AEI/10.13039/501100011033, ERDF/EU, ``European Union NextGenerationEU/PRTR", projects PID2022-142604OB-C21, PID2022-139519OB-C21, PID2023-147706NB-I00, PID2023-148896NB-I00, PID2023-146983NB-I00, PID2023-147115NB-I00, PID2023-148418NB-C41, PID2023-148418NB-C42, PID2023-148418NB-C43, PID2023-148418NB-C44, PID2024-158190NB-C22, RYC2021-033305-I, RYC2024-048719-I, CNS2023-144781, CNS2024-154769 and Plan de Ciencia, Tecnolog{\'i}a e Innovaci{\'o}n de Asturias, Spain; the Chulalongkorn Academic into Its 2nd Century Project Advancement Project, the National Science, Research and Innovation Fund program IND\_FF\_68\_369\_2300\_097, and the Program Management Unit for Human Resources \& Institutional Development, Research and Innovation, grant B39G680009 (Thailand); the Eric \& Wendy Schmidt Fund for Strategic Innovation through the CERN Next Generation Triggers project under grant agreement number SIF-2023-004; the Kavli Foundation; the Nvidia Corporation; the SuperMicro Corporation; the Welch Foundation, contract C-1845; and the Weston Havens Foundation (USA).
\end{acknowledgments}\section*{Data availability} Release and preservation of data used by the CMS Collaboration as the basis for publications is guided by the  \href{https://doi.org/10.7483/OPENDATA.CMS.1BNU.8V1W}{CMS data preservation, re-use and open access policy}.

\bibliography{auto_generated}
\cleardoublepage \appendix\section{The CMS Collaboration \label{app:collab}}\begin{sloppypar}\hyphenpenalty=5000\widowpenalty=500\clubpenalty=5000\cmsinstitute{Yerevan Physics Institute, Yerevan, Armenia}
{\tolerance=6000
A.~Gevorgyan\cmsorcid{0000-0003-2751-9489}, A.~Hayrapetyan, V.~Makarenko\cmsorcid{0000-0002-8406-8605}, A.~Tumasyan\cmsAuthorMark{1}\cmsorcid{0009-0000-0684-6742}
\par}
\cmsinstitute{Marietta Blau Institute for Particle Physics, Vienna, Austria}
{\tolerance=6000
W.~Adam\cmsorcid{0000-0001-9099-4341}, L.~Benato\cmsorcid{0000-0001-5135-7489}, T.~Bergauer\cmsorcid{0000-0002-5786-0293}, M.~Dragicevic\cmsorcid{0000-0003-1967-6783}, S.~Gundacker\cmsorcid{0000-0003-2087-3266}, A.K.~Guven\cmsorcid{0009-0004-5670-5138}, P.S.~Hussain\cmsorcid{0000-0002-4825-5278}, M.~Jeitler\cmsAuthorMark{2}\cmsorcid{0000-0002-5141-9560}, N.~Krammer\cmsorcid{0000-0002-0548-0985}, A.~Li\cmsorcid{0000-0002-4547-116X}, D.~Liko\cmsorcid{0000-0002-3380-473X}, M.~Matthewman, J.~Schieck\cmsAuthorMark{2}\cmsorcid{0000-0002-1058-8093}, R.~Sch\"{o}fbeck\cmsAuthorMark{2}\cmsorcid{0000-0002-2332-8784}, M.~Shooshtari\cmsorcid{0009-0004-8882-4887}, M.~Sonawane\cmsorcid{0000-0003-0510-7010}, N.~Van~Den~Bossche\cmsorcid{0000-0003-2973-4991}, W.~Waltenberger\cmsorcid{0000-0002-6215-7228}, C.-E.~Wulz\cmsAuthorMark{2}\cmsorcid{0000-0001-9226-5812}
% ...
\par}
\vskip\cmsinstskip
\dag:~Deceased\\
$^{1}$Also at Yerevan State University, Yerevan, Armenia\\
$^{2}$Also at TU Wien, Vienna, Austria\\
% ...
\end{sloppypar}
\end{document}